\documentclass[times,twocolumn,final,nopreprintline]{elsarticle}

\usepackage{framed,multirow}

\usepackage{amssymb}
\usepackage{latexsym}

\usepackage{url}
\usepackage{xcolor}

\usepackage{hyperref}
\usepackage{graphicx}
\usepackage{amsmath}
\usepackage{amssymb}
\usepackage{tabularx,longtable,multirow}
\usepackage{color}
\usepackage{xcolor}
\usepackage{tabularx}
\usepackage{supertabular}
\usepackage{epsfig}
\usepackage{graphicx}
\usepackage{amssymb}
\usepackage{colortbl}
\usepackage{relsize}
\usepackage{multirow}
\usepackage{float}

\usepackage[nolist]{acronym}
\usepackage[noend]{algpseudocode}
\usepackage[ruled,commentsnumbered]{algorithm2e}
\usepackage{booktabs}

\usepackage{makecell}
\usepackage{hhline}
\usepackage{array, spreadtab}
\usepackage{collcell,booktabs,siunitx}
\usepackage{xparse}
\usepackage{fp}
\usepackage{pgfplotstable}

\definecolor{yellow}{rgb}{0.8,0.6,0}
\definecolor{edited}{rgb}{0,0,0}
\definecolor{newcolor}{rgb}{.8,.349,.1}
\definecolor{ourrow}{RGB}{230,230,230}
\usepackage{animate}

\journal{Medical Image Analysis}  

\newcommand{\isnumber}[1]{%
  \ifnum\pdfmatch{^\d+(\.\d*)?$}{#1}=1 %
    1\relax
  \else
    0\relax
  \fi
}

\newcolumntype{Q}{>{\collectcell\rememberQ}c<{\endcollectcell}}
\newcolumntype{W}{>{\collectcell\rememberW}c<{\endcollectcell}}
\newcolumntype{E}{>{\collectcell\rememberE}c<{\endcollectcell}}
\newcolumntype{A}{>{\collectcell\computeA}c<{\endcollectcell}}

\newcommand{\rememberQ}[1]{\global\def\QValue{#1}\ignorespaces#1}
\newcommand{\rememberW}[1]{\global\def\WValue{#1}\ignorespaces#1}
\newcommand{\rememberE}[1]{\global\def\EValue{#1}\ignorespaces#1}

\ExplSyntaxOn
\NewDocumentCommand{\computeA}{} 
{
  \FPeval{\result}{round((\QValue+\WValue+\EValue)/3, 2)}
  \result
}
\ExplSyntaxOff

\newcolumntype{R}{>{\collectcell\rememberR}c<{\endcollectcell}}
\newcolumntype{T}{>{\collectcell\rememberT}c<{\endcollectcell}}
\newcolumntype{Y}{>{\collectcell\rememberY}c<{\endcollectcell}}
\newcolumntype{S}{>{\collectcell\computeS}c<{\endcollectcell}}

\newcommand{\rememberR}[1]{\global\def\RValue{#1}\ignorespaces#1}
\newcommand{\rememberT}[1]{\global\def\TValue{#1}\ignorespaces#1}
\newcommand{\rememberY}[1]{\global\def\YValue{#1}\ignorespaces#1}

\ExplSyntaxOn
\NewDocumentCommand{\computeS}{} 
{
  \FPeval{\result}{round((\RValue+\TValue+\YValue)/3, 2)}
  \result
}
\ExplSyntaxOff

\newcolumntype{U}{>{\collectcell\rememberU}c<{\endcollectcell}}
\newcolumntype{I}{>{\collectcell\rememberI}c<{\endcollectcell}}
\newcolumntype{O}{>{\collectcell\rememberO}c<{\endcollectcell}}
\newcolumntype{D}{>{\collectcell\computeD}c<{\endcollectcell}}

\newcommand{\rememberU}[1]{\global\def\LValue{#1}\ignorespaces#1}
\newcommand{\rememberI}[1]{\global\def\RValue{#1}\ignorespaces#1}
\newcommand{\rememberO}[1]{\global\def\MValue{#1}\ignorespaces#1}

\ExplSyntaxOn
\NewDocumentCommand{\computeD}{} 
{
  \FPeval{\result}{round((\UValue+\IValue+\OValue)/3, 2)}
  \result
}
\ExplSyntaxOff

\begin{document}



\begin{frontmatter}

\title{
Deformation-Recovery Diffusion Model (DRDM):\\ Instance Deformation for Image Manipulation and Synthesis}%



\author[1,2]{Jian-Qing Zheng\corref{cor1}\corref{cor2}}
\cortext[cor1]{These authors contributed equally to this work {}}
\cortext[cor2]{Corresponding author {}}
\ead{jianqing.zheng@\{ndm.ox.ac.uk; outlook.com\}}
\author[3,4]{Yuanhan Mo\corref{cor1}}
\author[3]{Yang Sun}
\author[3]{Jiahua Li}
\author[3]{Fuping Wu}
\author[5]{Ziyang Wang}
\author[1]{Tonia Vincent}
\author[3]{Bart{\l}omiej W. {Papie{\.z}}}

\address[1]{The Kennedy Institute of Rheumatology, University of Oxford, U.K.}
\address[2]{Chinese Academy of Medical Sciences Oxford Institute, University of Oxford, U.K.}
\address[3]{Big Data Institute, University of Oxford, U.K.}
\address[4]{\textcolor{edited}{MRC Laboratory of Medical Sciences, Imperial College London, U.K.}}
\address[5]{Department of Computer Science, University of Oxford, Oxford, U.K.
\\[12.5pt]
{\rm {\textbf{Project page}: \url{https://jianqingzheng.github.io/def_diff_rec/}}}
\vspace{-5pt}
}


\begin{abstract}

In medical imaging, diffusion models have shown great potential \textcolor{edited}{for} synthetic image generation. However, these \textcolor{edited}{approaches} often \textcolor{edited}{lack} interpretable correspondence between generated and \textcolor{edited}{real} images and \textcolor{edited}{can create anatomically implausible structures or} illusions. To address these \textcolor{edited}{limitations, we propose the Deformation-Recovery Diffusion Model (DRDM)}, a novel diffusion-based generative model \textcolor{edited}{that emphasizes morphological transformation} through deformation fields rather than direct image synthesis. 
\textcolor{edited}{DRDM introduces} a \textcolor{edited}{topology-preserving} deformation field generation \textcolor{edited}{strategy}, which randomly samples and integrates multi-scale Deformation Velocity Fields (DVFs). DRDM is trained to learn to recover \textcolor{edited}{unrealistic} deformation components, \textcolor{edited}{thus} restoring randomly deformed images to a realistic distribution. \textcolor{edited}{This formulation enables the generation of diverse yet anatomically plausible deformations that preserve structural integrity, thereby improving data augmentation and synthesis for downstream tasks} such as few-shot learning and image registration. 
\textcolor{edited}{Experiments on} cardiac \textcolor{edited}{Magnetic Resonance Imaging} and pulmonary \textcolor{edited}{Computed Tomography} show \textcolor{edited}{that} DRDM is capable of creating diverse, \textcolor{edited}{large-scale deformations, while maintaining anatomical plausibility of} deformation fields.
\textcolor{edited}{Additional evaluations on 2D image segmentation and 3D image registration tasks indicate notable performance gains, underscoring DRDM’s potential to enhance both image manipulation and generative modeling in medical imaging applications.}

\end{abstract}


\end{frontmatter}

\acrodef{AI}[AI]{Artificial Intelligence}
\acrodef{Attn}[Attn]{Attention-based}
\acrodef{ASD}[ASD]{Average Surface Distance}
\acrodef{BCE}[BCE]{Binary Cross Entropy}
\acrodef{Cas}[Cas]{Cascaded}
\acrodef{CT}[CT]{Computer Tomography}
\acrodef{C2F}[C2F]{Coarse-to-fine}
\acrodef{CPU}[CPU]{Central Processing Unit}
\acrodef{DDIM}[DDIM]{Denoising Diffusion Implicit Model}
\acrodef{DDPM}[DDPM]{Denoising Diffusion Probabilistic Model}
\acrodef{DDF}[DDF]{Dense Displacement Field}
\acrodef{DVF}[DVF]{Dense Velocity Field}
\acrodef{DRDM}[DRDM]{Deformation-Recovery Diffusion model}
\acrodef{DoF}[DoF]{Degree of Freedom}
\acrodef{DPRn}[DPRn]{Dual-stream Pyramidal Registration network}
\acrodef{DR}[DR]{Direct Regression}
\acrodef{DSC}[DSC]{Dice Similarity Coefficient}
\acrodef{FCNs}[FCNs]{Fully Convolution Networks}
\acrodef{FP}[FP]{Feature Pyramid}
\acrodef{GAN}[GAN]{Generative Adversarial Network}
\acrodef{GPU}[GPU]{Graphics Processing Unit}
\acrodef{HD}[HD]{Hausdorff Distance}
\acrodef{Iter}[Iter]{Iterative optimization}
\acrodef{LDM}[LDM]{Latent Diffusion Model}
\acrodef{M-H}[M-H]{Multi-Head}
\acrodef{MRBS}[MRBS]{multiple-resolution B-spline}
\acrodef{MRI}[MRI]{Magnetic Resonance Imaging}
\acrodef{MS}[MS]{Motion-Separable}
\acrodef{MSE}[MSE]{Mean Square Error}
\acrodef{prec}[prec]{precision}
\acrodef{RA}[RA]{Residual Aligner}
\acrodef{RAN}[RAN]{Residual Aligner-based Network}
\acrodef{RCn}[RCn]{Recursive Cascaded network}
\acrodef{ReLU}[ReLU]{Rectified Linear Unit}
\acrodef{sns}[sns]{sensitivity}
\acrodef{STN}[STN]{Spatial Transformer Network}
\acrodef{VAE}[VAE]{Variational Autoencoder}
\acrodef{VM}[VM]{Voxelmorph}

\begin{figure*}[!h]
\begin{center}
\includegraphics[width=0.7\linewidth]{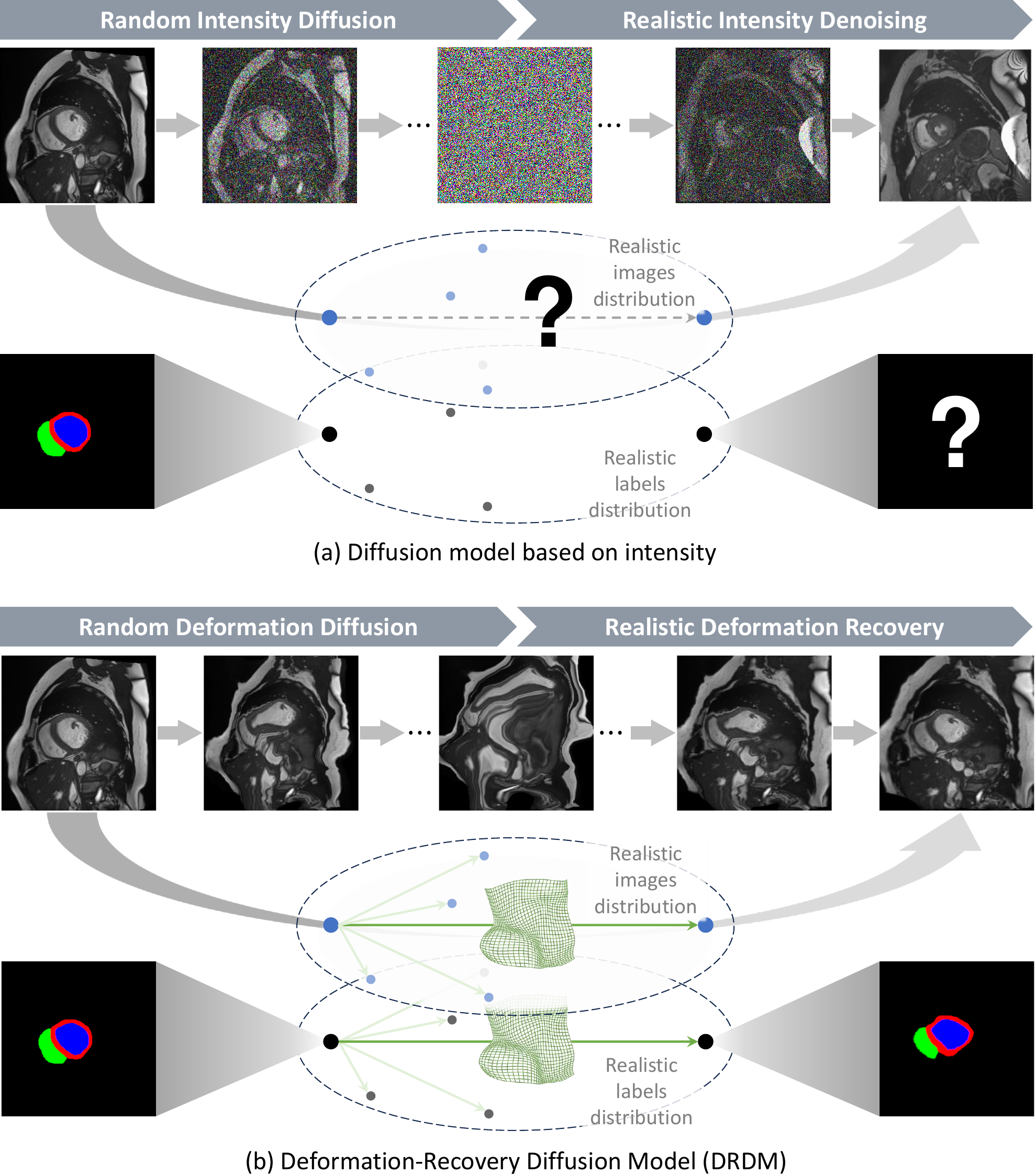}
\end{center}
\vspace{-.5em}
\caption{(a) \textcolor{edited}{Intensity-based diffusion models} can synthesize \textcolor{edited}{visually} realistic images, but \textcolor{edited}{lack an explicit} relationship with existing real subjects and thus unknown label \textcolor{edited}{relationship}; (b) \textcolor{edited}{In contrast, the proposed Deformation-Recovery diffusion model (DRDM) applies generated deformation fields to real images, representing anatomical variations.} \textcolor{edited}{These deformations can also be propagated to pixel-wise labels, thus enhancing the utility of the generated data for} downstream tasks.}
\label{fig:differ_diff}
\end{figure*}

\section{Introduction}
\label{sec:intro}

Image synthesis, a \textcolor{edited}{rapidly advancing} domain within \ac{AI}, has been revolutionized by the advent of deep learning technologies \citep{zhan2023multimodal}. It involves generating new images \textcolor{edited}{either from existing data or from random noise}, guided by specific \textcolor{edited}{structural, appearance, semantic, or statistical constraints}. Deep learning, with its ability to learn hierarchical representations \textcolor{edited}{\citep{lecun2015deep}}, has become the cornerstone of advancements in image synthesis, enabling applications that range from artistic image generation to the creation of realistic training data for machine learning models.

The heart of image synthesis via deep learning \textcolor{edited}{stems from the neural networks' ability to model} and manipulate complex data distributions. Generative models, particularly \acp{VAE} \citep{kingma2013auto} and \acp{GAN} \citep{goodfellow2014generative}, have emerged as powerful tools for this purpose. \acp{VAE} \textcolor{edited}{learn a compact} latent space representation, enabling the generation of new images by sampling from this space. \acp{GAN}, \textcolor{edited}{in contrast}, consist of a generator that synthesizes images and a discriminator that \textcolor{edited}{evaluates their realism}. The training process continues until the system reaches a Nash equilibrium, where the generator produces realistic images that the discriminator can no longer distinguish from real ones \citep{goodfellow2014generative}. 

Recently, intensity-based diffusion models, specifically \acp{DDPM} \citep{ho2020denoising}, have \textcolor{edited}{achieved state-of-the-art performance in image generation across various computer-vision tasks}. These models \textcolor{edited}{can} generate high-fidelity data and exhibit \textcolor{edited}{desirable} properties such as scalability and \textcolor{edited}{stable training}. \textcolor{edited}{Furthermore, latent diffusion models \citep{rombach2022high} extend \ac{DDPM} by operating in feature space, enabling multimodal data conditioning e.g. text-guided image generation.}

In medical imaging, diffusion models have been utilized for \textcolor{edited}{diverse tasks including} synthetic medical image generation \textcolor{edited}{\citep{pinaya2022brain,ju2024vm,du2023arsdm,bradbury2024paired}}, biomarker quantification \citep{gong2023diffusion}, anomaly detection \cite{li2023fast,bercea2023mask,liang2023modality}, image segmentation \textcolor{edited}{\citep{graf2023denoising,liang2023modality,zhang2024diffboost,zhang2025diffuseg}} and image registration \citep{qin2023fsdiffreg, gao2023diffusing}. These methods are capable of generating highly lifelike images but still suffer from potential issues such as producing visually plausible yet unrealistic artifacts and the inability of generated images to establish meaningful and interpretable relationships with pre-existing images, as illustrated in Figure~\ref{fig:differ_diff}(a). This limitation hinders their applicability in tasks, such as image segmentation, that require precise understanding and correlation with real data \citep{kazerouni2023diffusion,deo2025}.

Generating deformation fields rather than image intensities can address this issue by focusing on anatomical changes. Several previous works \textcolor{edited}{\citep{kim2022diffusemorph, kim2022diffusion, starck2024diff,wu2025igg,wang2025generating}} have attempted to generate deformation fields using image registration frameworks combined with diffusion models. However, they still employ diffusion-denoising approaches based on intensities \citep{kim2022diffusemorph, kim2022diffusion} or \textcolor{edited}{latent-feature} \citep{starck2024diff,wu2025igg,wang2025generating}, depending on registration frameworks to guide and constrain the rationality of the generated deformations. Consequently, the \textcolor{edited}{generated deformations are often limited to interpolating transformations between image pairs} \textcolor{edited}{\citep{kim2022diffusemorph, kim2022diffusion,wu2025igg}} or \textcolor{edited}{atlas-based deformation \citep{starck2024diff,wang2025generating}}, which restricts the ability to generate diverse and instance-specific morphological variations.

\begin{figure*}[!h]
\begin{center}
\includegraphics[width=0.9\linewidth]{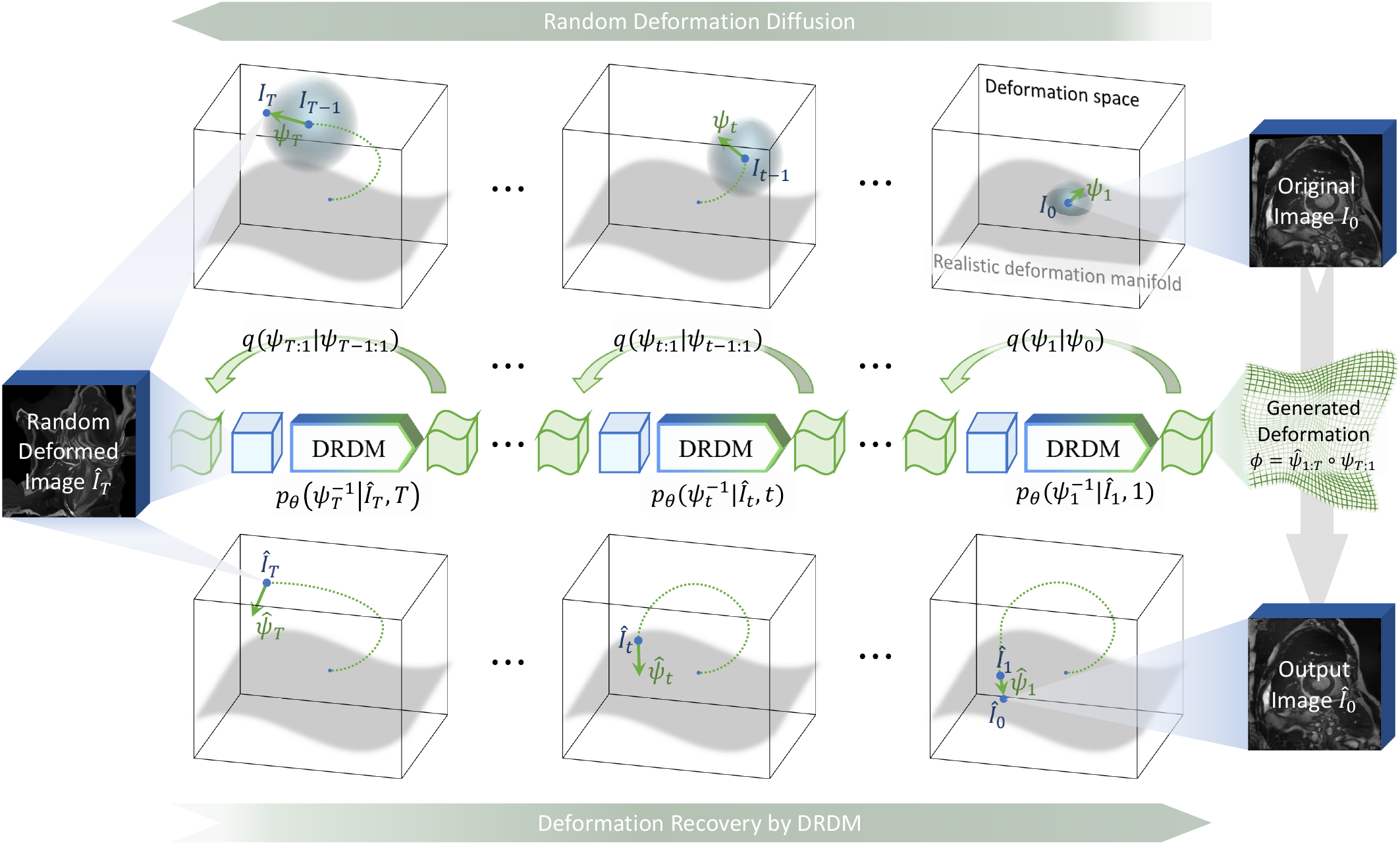}
\end{center}
\vspace{-.5em}
\caption{\textcolor{edited}{Overview of the DRDM framework. The model comprises two stages: (i) a deformation diffusion process that applies random deformations within the deformation space, and (ii) a deformation recovery process that recursively estimates and refines deformation field to generate a anatomically realistic image within the learned deformation manifold.}}
\label{fig:DRDM}
\end{figure*}

The noise added \textcolor{edited}{to image intensities} in the existing diffusion models \citep{sohl2015deep,ho2020denoising,song2020denoising} is \textcolor{edited}{independently distributed across pixels and typically follows} a normal distribution.
However, our \textcolor{edited}{objective} is to deform an \textcolor{edited}{existing} image rather than generate a new one, and \textcolor{edited}{deformation vectors across anatomical regions are inherently correlated}. 
Therefore, \textcolor{edited}{it is necessary to model deformation fields explicitly and design a diffusion process that reflects their realistic, spatially dependent distribution}. 

In \textcolor{edited}{this work}, we propose a novel diffusion generative model, \textcolor{edited}{centred on deformation rather than intensity, termed the \ac{DRDM}}.
\textcolor{edited}{\ac{DRDM} represents} a deformation-centric \textcolor{edited}{analogue} of the \textcolor{edited}{conventional} noising and denoising process, \textcolor{edited}{designed} to achieve realistic and diverse anatomical \textcolor{edited}{variations} as shown in Figure~\ref{fig:differ_diff}(b). As illustrated in Figure~\ref{fig:DRDM}, the framework includes \textcolor{edited}{two stages}: a random deformation diffusion \textcolor{edited}{process}, followed by realistic deformation recovery, \textcolor{edited}{which collectively enable} the generation of diverse deformations for individual images.
Our main contributions are \textcolor{edited}{summarized} as follows.
\begin{itemize}
    \item \textbf{Instance-specific deformation synthesis}: 
    {To the best of our knowledge, this is the first study to explore diverse deformation generation for individual images without \textcolor{edited}{utilizing any atlas/reference image or relying on population-level \textcolor{edited}{structural distributions \citep{he2024statistical,starck2024diff,wu2025igg,wang2025generating}}}};
    \item \textbf{Deformation Diffusion \textcolor{edited}{framework}}: We propose a novel diffusion model method based on deformation diffusion and recovery \textcolor{edited}{in contrast to intensity-based} diffusion \citep{kim2022diffusemorph,kim2022diffusion} or latent-feature diffusion \textcolor{edited}{\citep{starck2024diff,wu2025igg,wang2025generating}} based on registration frameworks;
    \item \textbf{Multi-scale random \ac{DVF} sampling and integrating}: 
    \textcolor{edited}{We propose a method for} multi-scale random \ac{DVF} sampling and integration to \textcolor{edited}{generate} deformation fields with physically possible distributions for \textcolor{edited}{training} \ac{DRDM};
    \item \textbf{Training from scratch without annotation}: 
    \textcolor{edited}{\ac{DRDM} is trained entirely from unlabelled images, without requiring human annotations or auxiliary (registration or optical/scene flow) model/framework};
    \item \textbf{Data augmentation for few-shot learning}:
    The deformation fields generated by \ac{DRDM} \textcolor{edited}{can be applied} on both image and \textcolor{edited}{corresponding} pixel-level segmentation, to augment morphological information \textcolor{edited}{while preserving anatomical topology, thus enabling data augmentation} for few-shot learning tasks;
    \item \textbf{Synthetic training for image registration}:
    {The synthetic deformations created by \ac{DRDM} can be used to train \textcolor{edited}{image registration models without the need for external annotations}};
    \item \textbf{Improved performance on downstream tasks}:
    The experimental results show that data augmentation \textcolor{edited}{and} synthesis by \ac{DRDM} improve performance in the downstream tasks, including segmentation and registration. \textcolor{edited}{Specifically, the \ac{DRDM}-augmented segmentation model outperforms the BigAug baselline \citep{BigAug}, and the DRDM-trained registration model surpasses prior synthetic training approaches \citep{eppenhof2018pulmonary}, validating the anatomical plausibility and utility of the learned deformation fields.}
\end{itemize}

The remainder of this paper is organized as follows. 
\textcolor{edited}{Section~\ref{sec:framework_design} presents the design of the \ac{DRDM} framework. Section~\ref{sec:img_syn} details the experimental setup and describes the generation of images and deformation fields. The applications of \ac{DRDM} to few-shot image segmentation and synthetic registration training are discussed in Sections~\ref{sec:aug_seg} and~\ref{sec:syn_reg}, respectively. Related work is reviewed in Section~\ref{sec:related_works}, and concluding remarks are provided in Section~\ref{sec:discuss_conclude}.}

\section{Framework design of DRDM}
\label{sec:framework_design}

The \textcolor{edited}{overall} framework of \textcolor{edited}{the proposed} \ac{DRDM} is \textcolor{edited}{illustrated} in Figure~\ref{fig:DRDM}.
The \textcolor{edited}{deformation field} generated \textcolor{edited}{by} \ac{DRDM}, \textcolor{edited}{represented as a dense displacement field \ac{DDF}}, is defined as a spatial transformation ${\phi}:\mathbb{R}^{H\times W \times D}\to\mathbb{R}^{H\times W \times D}$, \textcolor{edited}{where each element corresponds to}  a {displacement vector} denoted by $\phi[\textbf{\textit{x}}]\in \mathbb{R}^{3}$ at the \textcolor{edited}{voxel} coordinate $\textbf{\textit{x}}\in\mathbb{Z}^{3}$ of an image $\textbf{\textit{I}}\in\mathbb{R}^{{H\times W \times D}}$, where $H, W, D \in \mathbb{Z}_+$ denote the image height, width, and thickness, respectively. 
\textcolor{edited}{To maintain notational simplicity, all formulations in the following sections are presented for the 3D case, though they can be readily adapted to 2D by removing the depth dimension.}

The generation of plausible \ac{DDF} $\phi$ by \ac{DRDM} can be \textcolor{edited}{formulated as a composition of random deformation diffusion and deformation recovery processes}:
\textcolor{edited}{
\begin{equation}
\label{eq:phi_enc_dec}
\begin{array}{rl}
\phi&= \hat{\phi}_{1:T} \circ \phi_{T:1} \\
    &=\underbrace{\hat{\psi}_{1}\circ\hat{\psi}_{2}\cdots \hat{\psi}_{T}}_{\rm deformation~recovery}\circ \underbrace{\psi_{T}\circ\psi_{T-1}\cdots \psi_{1}}_{\rm deformation~diffusion}
\end{array}
\end{equation}
}
\textcolor{edited}{The first part of Equation~\eqref{eq:phi_enc_dec}, random deformation diffusion}, as described in Section~\ref{sec:forward_proc}, is to generate a \ac{DDF} \textcolor{edited}{($\phi_{t:1}$)} through a fixed Markov process of random \ac{DVF} generation and integration of the \ac{DVF}s \textcolor{edited}{($\psi_{t},\cdots,\psi_{1}$)}:
\textcolor{edited}{
\begin{equation}
\label{eq:phi_enc}
\phi_{t:1}:=\psi_{t}\circ\psi_{t-1}\cdots\psi_{1} \sim q(\phi_{t:1}|~\phi_{t-1:1})  \\
\end{equation}
}
\textcolor{edited}{The second part of Equation~\eqref{eq:phi_enc_dec},} deformation recovery, as described in Section~\ref{sec:backward_proc}, is to estimate the recovering \ac{DDF} \textcolor{edited}{$\hat{\phi}_{t:T}$} with the inverse \ac{DVF} for each step ${\psi}^{-1}_{t}$ estimated as $\hat{\psi}_{t}$ recursively based on the input of the deformed image ${\textbf{\textit{I}}}_{t}$:
\textcolor{edited}{
\begin{equation}
\label{eq:phi_dec}
\left\{
\begin{array}{c}
\hat{\phi}_{t:T}:=\hat{\psi}_{t}\circ\hat{\psi}_{t+1}\cdots \hat{\psi}_{T}\\
\hat{\psi}_{t}\sim p(\psi_{t}^{-1}|~\hat{\textbf{\textit{I}}}_{t},t)  \\
\end{array}
\right.
\end{equation}
}
where $\textit{\textbf{I}}_{0}, \textit{\textbf{I}}_{t}$ denote the original image and the randomly deformed image respectively, $\hat{\textit{\textbf{I}}}_{0}$ and $\hat{\textit{\textbf{I}}}_{t}$ denote the synthesised image by \ac{DRDM} and the intermediate recovered images, respectively. 
\textcolor{edited}{These are calculated as}:
\textcolor{edited}{
\begin{equation}
\label{eq:deformed_img}
\left\{
\begin{array}{c}
{\textbf{\textit{I}}}_{t}:={\phi}_{t:1}(\textbf{\textit{I}}_{0})\\
\hat{\textbf{\textit{I}}}_{t}:=\langle\hat{\phi}_{t+1:T}\circ{\phi}_{T:1}\rangle(\textbf{\textit{I}}_{0})
\end{array}
\right.
\end{equation}
}
where $0<t\leq T$ denotes the deformation step in diffusion or recovery processing, $T$ denotes the total number of deformation steps for the diffusion and recovery process, and $\circ$ denotes the composition operation. 
\textcolor{edited}{This} is calculated by:
\begin{equation}
\label{eq:composite}
\langle\phi_1\circ \phi_2\rangle[x]:=\phi_2[\phi_1[x]+x]+\phi_1[x]
\end{equation}
to simulate the process of gradual deformation \citep{arsigny2006log,vercauteren2009diffeomorphic}.
\textcolor{edited}{
Unlike intensity-based diffusion models \citep{ho2020denoising,song2020denoising}, which add and normalise independent noise components across time steps, the \ac{DRDM} formulation explicitly models spatially correlated deformation compositions in the transformation manifold.}

The transformation of images by a given deformation field and the composition between two deformation fields are implemented based on the \ac{STN} \citep{jaderberg2015spatial}.


\begin{figure*}[!ht]
\begin{center}
\includegraphics[width=0.7\linewidth]{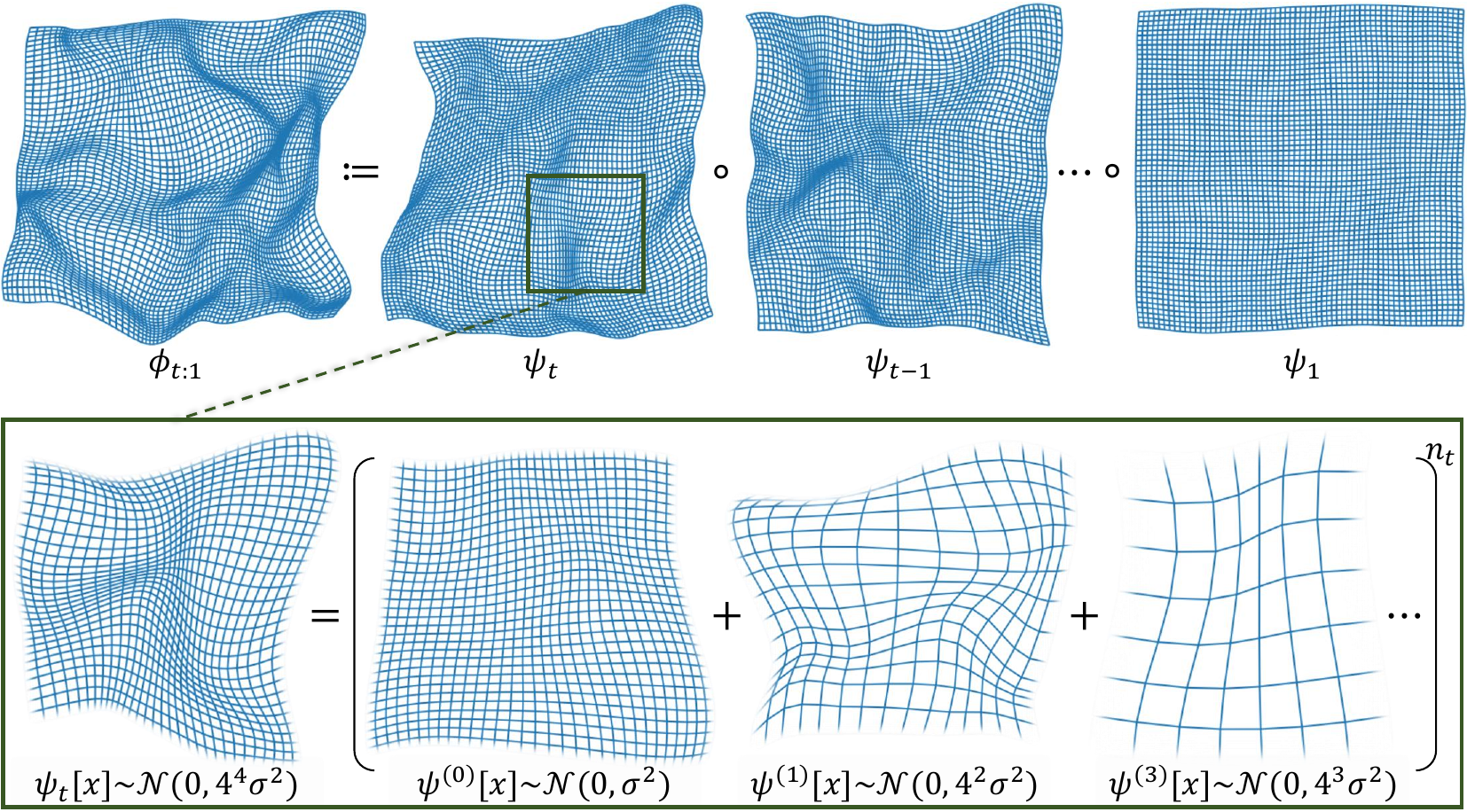}
\end{center}
\vspace{-.5em}
\caption{Illustration of the principle underlying multi-scale random DVF generation and integration in the deformation diffusion process, as detailed in Equation~\eqref{eq:phi_enc} and \eqref{eq:multi_scale_dvf_syn}. \textcolor{edited}{(edited according to Comment-2.3)}}
\label{fig:forward_proc}
\end{figure*}

\subsection{Forward process for random deformation diffusion}
\label{sec:forward_proc}
This section introduces the forward processing for random deformation diffusion as illustrated in Figure~\ref{fig:forward_proc}.
This process defines how physically plausible random deformations are generated within the \ac{DRDM} framework for medical imaging applications. To ensure anatomical realism, the underlying assumptions and deformation constraints are introduced in Section~\ref{sec:def_rule}. Based on these assumptions, Section~\ref{sec:rand_def_synth} describes the random generation of noisy deformation velocity fields (\acp{DVF}). The procedure for determining the corresponding noise levels is presented in Section~\ref{sec:noise_setting}, and the integration of \acp{DVF} into dense displacement fields (\acp{DDF}) is detailed in Section~\ref{sec:noisy_def_comp}.

\subsubsection{The nature of deformation}
\label{sec:def_rule}
As presented previously, the Gaussian noise applied to image intensities in conventional diffusion models \citep{sohl2015deep,ho2020denoising,song2020denoising} is independent for each pixel/voxel. In contrast, anatomical deformations exhibit spatial correlations, as neighboring regions of an organ typically move coherently.
To ensure anatomically plausible transformations during the forward deformation diffusion process, a set of physical and topological constraints is defined to regularize the random deformation generation:
\begin{enumerate}
    \item Randomness: The deformation vector of each position should yield \textcolor{edited}{a} normal distribution $\psi_t[x]\sim \mathcal{N}(0,\sigma_t^2)$;
    \item Local dependency: the deformation field of a continuum should be continuous and thus the stochastic regional discontinuity is bounded by $\Delta(\psi_t,\Delta x):=\psi_t[x+\Delta x]-\psi_t[x]\sim\mathcal{N}(0,\sigma'_t(\Delta x)^2)$, where $\sigma'_t(\Delta x_1)\geq\sigma'_t(\Delta x_2),~{\|\Delta x_1\|}_\infty> {\|\Delta x_2\|}_\infty$;
    \item \textcolor{edited}{Diffeomorphism}: \textcolor{edited}{the generated deformation field of a continuum should preserve anatomical topology}: $|{\textbf{\textit{J}}}|_{<0}<\epsilon$.
\end{enumerate}
where Chebyshev distance ${\|\cdot\|}_\infty$ is used \textcolor{edited}{to measure spatial neighbourhood relationships}, $|{\textbf{\textit{J}}}|_{<0}$ denotes \textcolor{edited}{the proportion of voxels with negative Jacobian determinant values of the deformation field}, $\epsilon$ denotes a small positive value to constrain the unrealistic deformation, $\sigma_t^2$ denotes the deformation variance of \ac{DVF} $\psi_{t}$ at the $t^{\rm th}$ time step, and ${\sigma'}_t^2$ denotes the deformation discontinuity variance of \ac{DVF} $\psi_{t}$.

These rules are primarily formulated for modeling the deformation of a single continuous structure. However, in cases involving discontinuous deformations across multiple organs or tissue interfaces, the situation becomes more complex, as previously discussed in \citep{papiez2014implicit, zheng2022residual}.

\subsubsection{Multi-scale random {DVF} generation}
\label{sec:rand_def_synth}

Following the deformation constraints described in Section~\ref{sec:def_rule}, a multi-scale random \ac{DVF} is synthesized at each time step by sampling from multiple Gaussian distributions at different spatial scales:
\begin{equation}
\label{eq:multi_scale_dvf_syn}
\left\{
\begin{array}{cc}
\psi=\psi^{(0)}+{\rm intrp}(\psi^{(1)})+\cdots+{\rm intrp}(\psi^{(m)}) \\
\psi^{(0)}\in \mathbb{R}^{3 \times h\times w \times d}, ~\psi^{(0)}[x]\sim \mathcal{N}(0,\sigma^2)\\
\psi^{(1)}\in \mathbb{R}^{3 \times (h/2) \times (w/2) \times (d/2)}, ~\psi^{(1)}[x]\sim \mathcal{N}(0,(2\sigma)^2)\\
\cdots\\
\psi^{(m)}\in \mathbb{R}^{3 \times (h/2^m)\times (w/2^m) \times (d/2^m)}, ~\psi^{(m)}[x]\sim \mathcal{N}(0,(2^m\sigma)^2)
\end{array}
\right.
\end{equation}
where ${\rm intrp(\cdot)}$ denotes interpolation of the input image/\ac{DDF}/\ac{DVF} to the spatial resolution of $h\times w\times d$. \textcolor{edited}{The comoponents $\psi^{(0)},\psi^{(1)},\cdots \psi^{(m)}$ represent the independent \ac{DVF} samples at different scales, namely the original scale, the first-order half-down-sampled scale, $\cdots$ and up to the $m^{\rm th}$-order half-down-sampled scale.} 
\textcolor{edited}{Under these definitions, the first constraint (randomness) described in Section~\ref{sec:def_rule} is satisfied when}:
\begin{equation}
\label{eq:rule_1}
\sigma_t^2 \approx \frac{4^{m+1}-1}{3}\sigma^2n_t
\end{equation}
\textcolor{edited}{and the second constraint (local dependency) is satisfied when}:
\begin{equation}
\label{eq:rule_2}
{\sigma'_t}^2 \approx2 n_t \sigma^2\sum_{i=0}^{m}{\min({\|\Delta x\|}_\infty,2^i)^2}
\end{equation}
where $\sigma^2$ denotes the minimum unit of \ac{DVF} variance, $n_t$ denotes the noise scale \textcolor{edited}{factor} for each \textcolor{edited}{diffusion} time step, as described in Section~\ref{sec:noise_setting}.

\subsubsection{Noise scale of the random deformation field}
\label{sec:noise_setting}
To ensure the \textcolor{edited}{diffeomorphism} of the generated deformation fields, the \ac{DDF} is modeled as a pseudo flow, which can be differentiated into a \ac{DVF} at each time step, following the continuum flow \textcolor{edited}{formulation} \citep{christensen1996deformable}.
For sampling of \ac{DVF} with varying \textcolor{edited}{variance of magnitude} at different time steps, an initial \ac{DVF} is first sampled with a small fixed variance, and then integrated recursively to a larger \ac{DVF} \textcolor{edited}{where the number of integration iterations $n_t$ controls the overall deformation scale in the forward process}.
The integrating recursion number \textcolor{edited}{is} used to control the magnitude of the random deformation field in the forward process.
The noise scaling level is set to increase with the increasing time step $t$, \textcolor{edited}{expressed as}:
\begin{equation}
\label{eq:noise_t}
n_t:={t}^{\alpha}/\beta
\end{equation}
where $\alpha$ and $\beta$ denote the parameters \textcolor{edited}{that control the rate of noise-level growth across diffusion steps}.

\subsubsection{Deformation diffusion by integrating DVF to DDF}
\label{sec:noisy_def_comp}

As described in Equation~\eqref{eq:phi_enc}, the \textcolor{edited}{generated} \ac{DVF} in Section~\ref{sec:rand_def_synth} isintegrated to \ac{DDF} \textcolor{edited}{$\phi_{t:1}$} by composing \textcolor{edited}{the sequence of deformation velocity fields} $\psi_t,~\psi_{t-1},\cdots\psi_1$. Thus the random deformation field \textcolor{edited}{$\phi_{t:1}[x]\sim\mathcal{N}(0,\sigma_{t:1}^2)$} can be sampled \textcolor{edited}{as}:
\begin{equation}
\label{eq:sigma_t_1}
\sigma_{t:1}^2:=\sum_{i=1}^{t}\sigma_t^2\approx \frac{4^{m+1}-1}{3}n_{t:1}\sigma^2
\end{equation}
where $n_{t:1}$ denotes the integrated noise scale \textcolor{edited}{factor, defined as}:
\begin{equation}
\label{eq:noise_t_1}
n_{t:1}:=\int_{1}^{t}{n_\tau {\rm d}\tau}\approx \frac{t^{\alpha+1}}{(\alpha+1)\beta}
\end{equation}

\begin{figure*}[!ht]
\begin{center}
\includegraphics[width=0.8\linewidth]{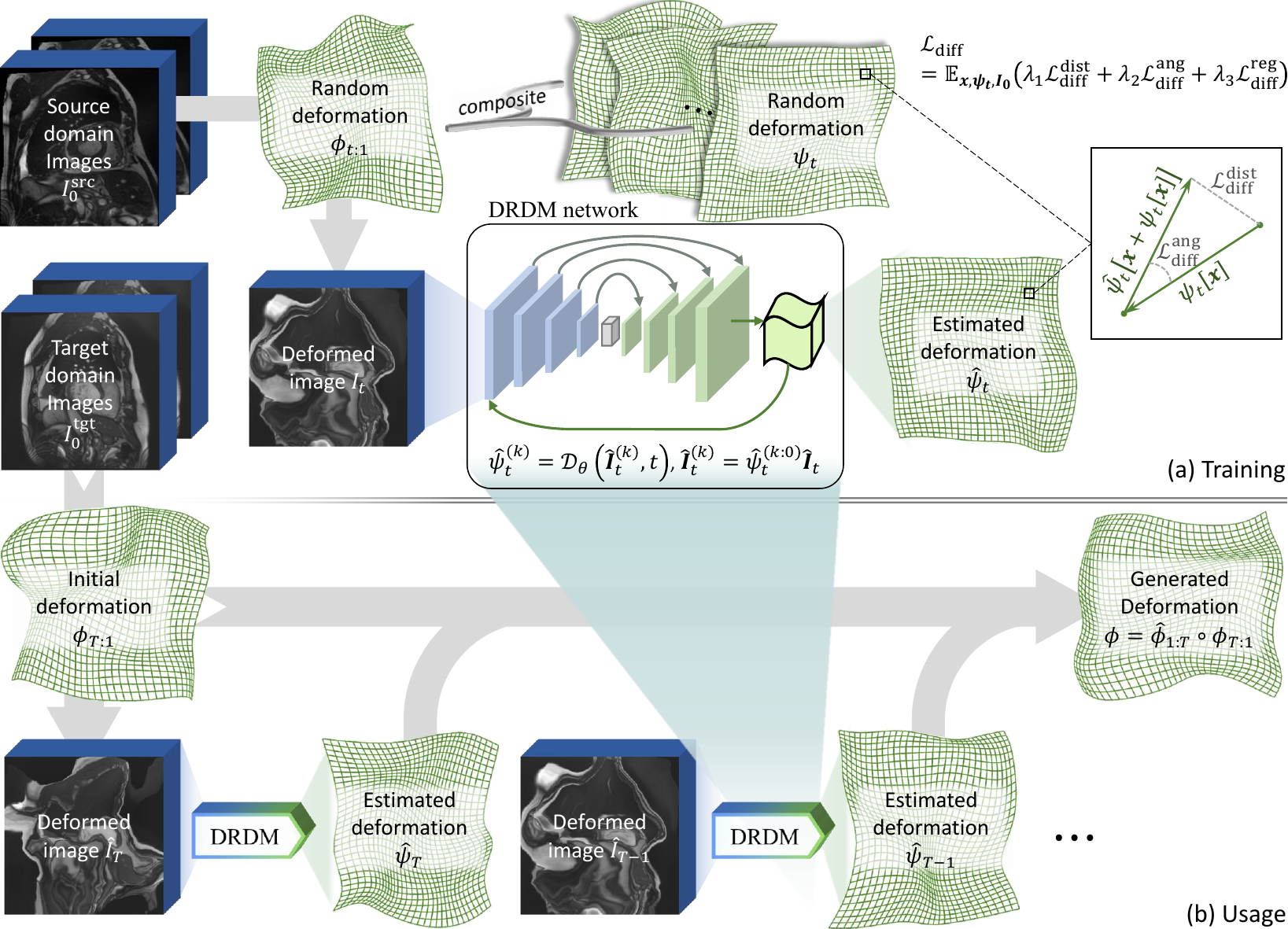}
\end{center}
\vspace{-.5em}
\caption{(a) The DRDM is trained \textcolor{edited}{at} each time step using distance- and angle-based loss function, \textcolor{edited}{as described in} Algorithm~\ref{algo:train_drdm}. (b) \textcolor{edited}{During inference,} deformation fields are \textcolor{edited}{generated by the DRDM} with varying time step and integrated to \textcolor{edited}{produce} the final deformation $\phi$ according to Algorithm~\ref{algo:infer_drdm}. (\textcolor{edited}{edited according to Comment-2.3})}
\label{fig:net_design}
\end{figure*}

\subsection{Reverse process for deformation recovery}
\label{sec:backward_proc}

\textcolor{edited}{Unlike} pixel-wise intensity \textcolor{edited}{prediction} by \ac{DDPM} or \ac{DDIM} \citep{ho2020denoising,song2020denoising}, the \textcolor{edited}{proposed \ac{DRDM} is designed} to estimate a deformation field. Figure~\ref{fig:net_design} \textcolor{edited}{illustrates} the training and usage pipeline of the network for \ac{DRDM}.
The \textcolor{edited}{architecture of \ac{DRDM} network is presented in Section~\ref{sec:net_design},
the training strategy} is described in Section~\ref{sec:net_optimize}, and \textcolor{edited}{the application of the trained \ac{DRDM} for instance-specific image deformation is described in} Section~\ref{sec:deform_synth_using_net}.

\subsubsection{Recursive network design for DRDM}
\label{sec:net_design}

As described in Equation~\eqref{eq:phi_dec}, the \ac{DVF} \textcolor{edited}{used for deformation recovery} is estimated and sampled by \ac{DRDM} $\mathcal{D}_\theta$ based on $\hat{\psi}_t \sim p(\psi_{t}^{-1}|~\hat{\textbf{\textit{I}}}_{t},t)$ with the input image $\hat{\textbf{\textit{I}}}_t$ \textcolor{edited}{at} the time step $t$:
\begin{equation}
\label{eq:net_est_dvf}
\left\{
\begin{array}{cl}
\hat{\psi}_t^{(0)}:\hat{\textbf{\textit{I}}}_t \mapsto \hat{\textbf{\textit{I}}}_t&\\
\hat{\psi}_t^{(k)}=\mathcal{D}_\theta(\hat{\textbf{\textit{I}}}_t^{(k)},~t) \\
\hat{\textbf{\textit{I}}}_t^{(k)}=\langle\hat{\psi}_t^{(k-1)}\circ\hat{\psi}_t^{(k-2)}\cdots\hat{\psi}_t^{(0)}\rangle(\hat{\textbf{\textit{I}}}_t)\\
\hat{\psi}_t=\hat{\psi}_t^{(K)}\circ\hat{\psi}_t^{(K-1)}\cdots\hat{\psi}_t^{(1)}&
\end{array}\right.
\end{equation}
where \ac{DRDM} $\mathcal{D}_\theta$ estimates a set of \ac{DVF} $\hat{\psi}_t^{(k)}$ \textcolor{edited}{though the} internal recursion $1\leq k \leq K$ and integrates them to regress the inverse \ac{DVF} $\psi_t^{-1}$. 

The U-Net \textcolor{edited}{architecture} \citep{ronneberger2015u} is adapted into a recursive structure \textcolor{edited}{and combined with} Atrous II blocks \citep{zhou2020acnn} to \textcolor{edited}{enlarge the receptive field, thereby improving the network’s ability to capture spatial context} \citep{islam2019much}. The detailed network architecture is \textcolor{edited}{provided} in \ref{sec:net_detail_drdm}. 

The internal recursion is designed to ensure that the network can adapt to \textcolor{edited}{each} input deformed image in a single-step training strategy, and \textcolor{edited}{the number of internal iterations $K$ is set to 2 following} \citep{zheng2022recursive}.

\textcolor{edited}{
Notably, in both Equations~\eqref{eq:deformed_img} and \eqref{eq:net_est_dvf}, multiple deformation fields are applied via compositional warping, rather repeatedly deforming the image ($\textbf{\textit{I}}_0$ and $\textbf{\textit{I}}_t^{(0)}$) itself, which prevents blurring and preserves fine anatomical structures.
}

\begin{algorithm}[th]
    \label{algo:train_drdm}
    \caption{Training \ac{DRDM}}
    \KwIn{Training set of source domain images $\mathbf{D}^{\rm src}\subset{\mathbb{R}^{H\times W\times D}}$}
    \KwOut{\ac{DRDM} weights $\theta$}
    \vspace{-0.5em}
    \hrulefill\\
    \nl initialize the DRDM parameters $\theta$\;
    \nl \While{${\mathcal{L}_{\rm diff}}$ not converge}{
    \tcp{randomly sample the data}
    \nl sample the original images: $\textbf{\textit{I}}_0\in \mathbf{D}^{\rm src}$\;
    \nl sample time steps: $t\sim\mathcal{U}(0,T)\cap\mathbb{Z}$\;
    \nl sample random DVFs $\psi_t$ and DDFs $\phi_{t:1}$ according to \eqref{eq:multi_scale_dvf_syn}, \eqref{eq:rule_1} and \eqref{eq:sigma_t_1}\;
    \tcp{compute the prediction and the loss}
    \nl deform original images from ${\textbf{\textit{I}}}_0$ to ${\textbf{\textit{I}}}_t$ using \eqref{eq:deformed_img}\;
    \nl use DRDM $\mathcal{D}_\theta$ to estimate recovering deformation $\hat{\psi}_t$ via \eqref{eq:net_est_dvf}\;
    \nl Update gradient descent step $\nabla_\theta{\mathcal{L}_{\rm diff}}$ via \eqref{eq:loss_DRDM}\;
    }
    \nl return model weights $\theta$.
\end{algorithm}

\subsubsection{Network optimizing for {DRDM}}
\label{sec:net_optimize}

The \ac{DRDM} $\mathcal{D}_{\theta}$ is trained \textcolor{edited}{on} randomly sampled time step $t\sim\mathcal{U}(0,T)\cap\mathbb{Z}$
\textcolor{edited}{where the trainable parameters $\theta$ are optimized by minimizing the following}:
\begin{equation}
\label{eq:opt_DRDM}
\min_{\theta}{\{\mathcal{L}_{\rm diff}({\psi_t},\hat{\psi}_t)\}}
\end{equation}
where the \textcolor{edited}{total} loss function $\mathcal{L}_{\rm diff}$ is \textcolor{edited}{defined} by:
\begin{equation}
\label{eq:loss_DRDM}
\left\{
\begin{array}{l}
\mathcal{L}_{\rm diff}:=\mathbb{E}_{x,\psi_t,\textbf{\textit{I}}_0}\big(\lambda_{1}\mathcal{L}_{\rm diff}^{\rm dist}+\lambda_{2}\mathcal{L}_{\rm diff}^{\rm ang}+\lambda_{3}\mathcal{L}_{\rm diff}^{\rm reg}\big)\\
\mathcal{L}_{\rm diff}^{\rm dist}:={\frac{\|\langle\psi_t\circ\hat{\psi}_t\rangle[x]\|_2}{\|\psi_t[x]\|_2+\epsilon}}\\
\mathcal{L}_{\rm diff}^{\rm ang}:=-{\frac{{{\psi}_t[x]}^{\top}\hat{\psi}_t[x+\psi_t[x]]}{\|{{\psi}_t[x]}\|_2 \|\psi_t(\hat{\psi}_t[x])\|_2+\epsilon}}\\
\mathcal{L}_{\rm diff}^{\rm reg}:={{\|\nabla_x{\hat{\psi}_t}[x]\|}_1 + {\rm relu}(-{\rm det}(\nabla_x\hat{\psi}_t[x]))}\\
\end{array}\right.
\end{equation}
\textcolor{edited}{Here,} ${\rm det}(\cdot)$ denotes the determinant of a matrix, $\nabla\hat{\psi}_t$ denotes the Jacobian matrix of the estimated \ac{DVF}.
\textcolor{edited}{The total} loss function for training the \ac{DRDM} model consists of three terms \textcolor{edited}{as follows}: 
(i) the distance error loss$\mathcal{L}_{\rm diff}^{\rm dist}$, \textcolor{edited}{which measures the magnitude difference between the true and estimated deformations;}
(ii) the angle error loss $\mathcal{L}_{\rm diff}^{\rm ang}$, \textcolor{edited}{which penalizes orientation discrepancies between the deformation vectors;} and
(iii) the regularization $\mathcal{L}_{\rm diff}^{\rm reg}$, \textcolor{edited}{which enforces spatial smoothness and penalizes non-diffeomorphic regions via the L1-norms and the negative determinant of the Jacobian}.
The relative importance of these components is controlled by the weighting factors $\lambda{1}$, $\lambda_{2}$, and $\lambda_{3}$.

As shown in Algorithm~\ref{algo:train_drdm}, the weights of \ac{DRDM} $\theta$ are \textcolor{edited}{optimized using} a set of training images from \textcolor{edited}{the} source domain ($\textbf{\textit{I}}_0\in\mathbf{D}^{\rm src}\subset{\mathbb{R}^{H\times W\times D}}$). 
\textcolor{edited}{The training procedure begins by initializing $\mathcal{D}_{\theta}$ and iteratively sampling random time steps $t$ from a uniform distribution. At each iteration, random \acp{DVF} ($\psi_t$) and \acp{DDF} ($\phi_{t:1}$) are generated according to the forward diffusion process. The original images $\textbf{\textit{I}}_0$ are then deformed into new states $\textbf{\textit{I}}_t$, and the network predicts the corresponding inverse deformation $\hat{\psi}_t$ required to recover the original image.}
The model parameters ($\theta$) are updated through gradient descent to minimize the loss $\mathcal{L}_{\rm diff}$, improving the model's deformation understanding and recovering capabilities. The training ends when the optimized weights ($\theta$) are finalized upon convergence of the loss function.

\begin{algorithm}[th]
    \label{algo:infer_drdm}
    \caption{Instance Deformation via \ac{DRDM}}
    \KwIn{Images for deformation $\textbf{\textit{I}}_0\in{\mathbb{R}^{H\times W\times D}}$}
    \KwOut{Generated \ac{DDF} $\phi$}
    \vspace{-0.5em}
    \hrulefill\\
    \nl import the DRDM parameters $\theta$ from Algorithm~\ref{algo:train_drdm}\;
    \nl set the deformation level $T'\leq T$\;
    \tcp{deformation diffusion process}
    \nl sample a random DDF $\phi_{T':1}$ using \eqref{eq:multi_scale_dvf_syn} and \eqref{eq:sigma_t_1}\;
    \nl set the initial DDF for deformation recovery: $\phi\leftarrow{\phi_{T':1}}$\;
    \nl deform original images from ${\textbf{\textit{I}}}_0$ to ${\textbf{\textit{I}}}_{T'}$ using \eqref{eq:deformed_img}\;
    \nl set the initial image for deformation recovery: $\hat{\textbf{\textit{I}}}_{T'}\leftarrow{\textbf{\textit{I}}}_{T'}$\;
    \tcp{deformation recovery process}
    \nl \For{$t=T',T'-1,\cdots,1$}{
    \nl use DRDM $\mathcal{D}_\theta$ to estimate recovering deformation $\hat{\psi}_t$ via \eqref{eq:net_est_dvf}\;
    \nl update the deformation according to \eqref{eq:phi_enc_dec}: $\phi\leftarrow \hat{\psi}_t\circ\phi$\;
    \nl deform original images from ${\textbf{\textit{I}}}_0$ to $\hat{\textbf{\textit{I}}}_{t-1}$ using \eqref{eq:deformed_img}\;    
    }
    \nl return the generated deformation $\phi$.
\end{algorithm}

\subsubsection{Instance deformation synthesis by DRDM}
\label{sec:deform_synth_using_net}

After \textcolor{edited}{training} the \ac{DRDM}, the deformation field \ac{DDF} $\phi$ is generated \textcolor{edited}{according to} Algorithm~\ref{algo:infer_drdm}, as shown in Figure~\ref{fig:net_design}(b).
The algorithm generates a \ac{DDF} through \textcolor{edited}{composing} a sequence of \acp{DVF} produced by the trained \ac{DRDM} $\mathcal{D}_\theta$. Starting \textcolor{edited}{from} an initial image from the target domain, represented as $\textbf{\textit{I}}_0\in\mathbf{D}^{\rm tgt}$, with the size of height $H$, width $W$, and depth $D$, the algorithm \textcolor{edited}{performs a series of steps to produce the final deformation}.

Initially, a deformation step level $T'$ \textcolor{edited}{is defined, not exceeding the maximum diffusion} step level $T$. 
\textcolor{edited}{A} random \ac{DDF} \textcolor{edited}{$\phi_{T':1}$} \textcolor{edited}{is then sampled using the} multi-scale \ac{DVF} synthesis Equation~\eqref{eq:multi_scale_dvf_syn} and \eqref{eq:sigma_t_1}. This sampled \ac{DDF} is set as the initial \ac{DDF} $\phi$ for the \textcolor{edited}{subsequent} deformation recovery process.

The \textcolor{edited}{input} image $\textbf{\textit{I}}_0$ \textcolor{edited}{is first deformed to produce} $\textbf{\textit{I}}_{T'}$, \textcolor{edited}{which serves as the} initial state for deformation recovery, $\hat{\textbf{\textit{I}}}_{T'}$. 
The \textcolor{edited}{recovery process proceeds through} reverse iteration from $t = T'$ down to 1. \textcolor{edited}{At} each iteration, the \ac{DRDM} network estimates a recovering deformation $\hat{\psi}_t$, \textcolor{edited}{which} is used to update $\phi$ \textcolor{edited}{by} integrating the current \textcolor{edited}{estimate} with the accumulated deformations from previous steps.

Each iteration not only updates \textcolor{edited}{both} the deformation field ${\phi}$ \textcolor{edited}{and the corresponding deformed image}. \textcolor{edited}{Specifically, the deformation is applied to the original image $\textbf{\textit{I}}_0$, generating a progressively updated image state $\hat{\textbf{\textit{I}}}_{t-1}$ that transitions smoothly} from ${\textbf{\textit{I}}}_{T’}$. The algorithm concludes by returning the fully \textcolor{edited}{integrated} \ac{DDF} $\phi$, representing the cumulative deformation applied to the original image to reach \textcolor{edited}{its final deformed configuration}.

\textcolor{edited}{This iterative estimation enables DRDM to model complex, non-linear anatomical variations, with the total number of deformation steps $T'$ controlling the overall deformation magnitude.}
 

\begin{figure}[!h]
\begin{center}
\includegraphics[width=\linewidth]{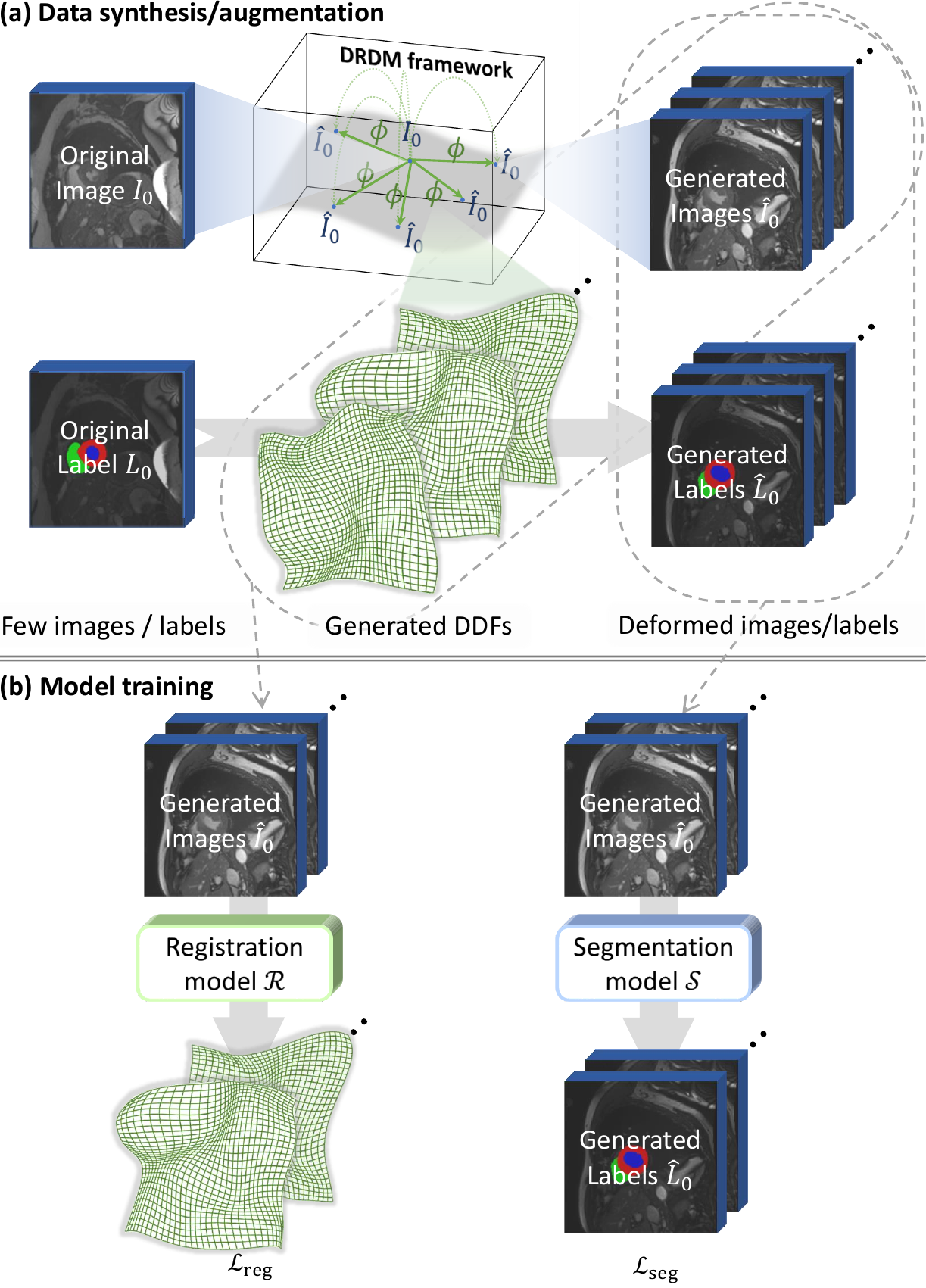}
\end{center}
\vspace{-.5em}
\caption{Image and deformation synthesis \textcolor{edited}{using the proposed} DRDM for few-shot-learning in image segmentation and image registration. (a) Diverse deformation fields, images, and corresponding labels are generated based on the input few images with labels, as described in Algorithm~\ref{algo:data_aug} and Algorithm~\ref{algo:data_syn}; (b) The \textcolor{edited}{synthesized} images and the corresponding labels are used to train a segmentation model, \textcolor{edited}{while the} generated images \textcolor{edited}{and the} corresponding DDFs are \textcolor{edited}{employed} to train a registration model.}
\label{fig:data_aug}
\end{figure}

\begin{figure*}[!h]
\begin{center}
\includegraphics[width=0.9\linewidth]{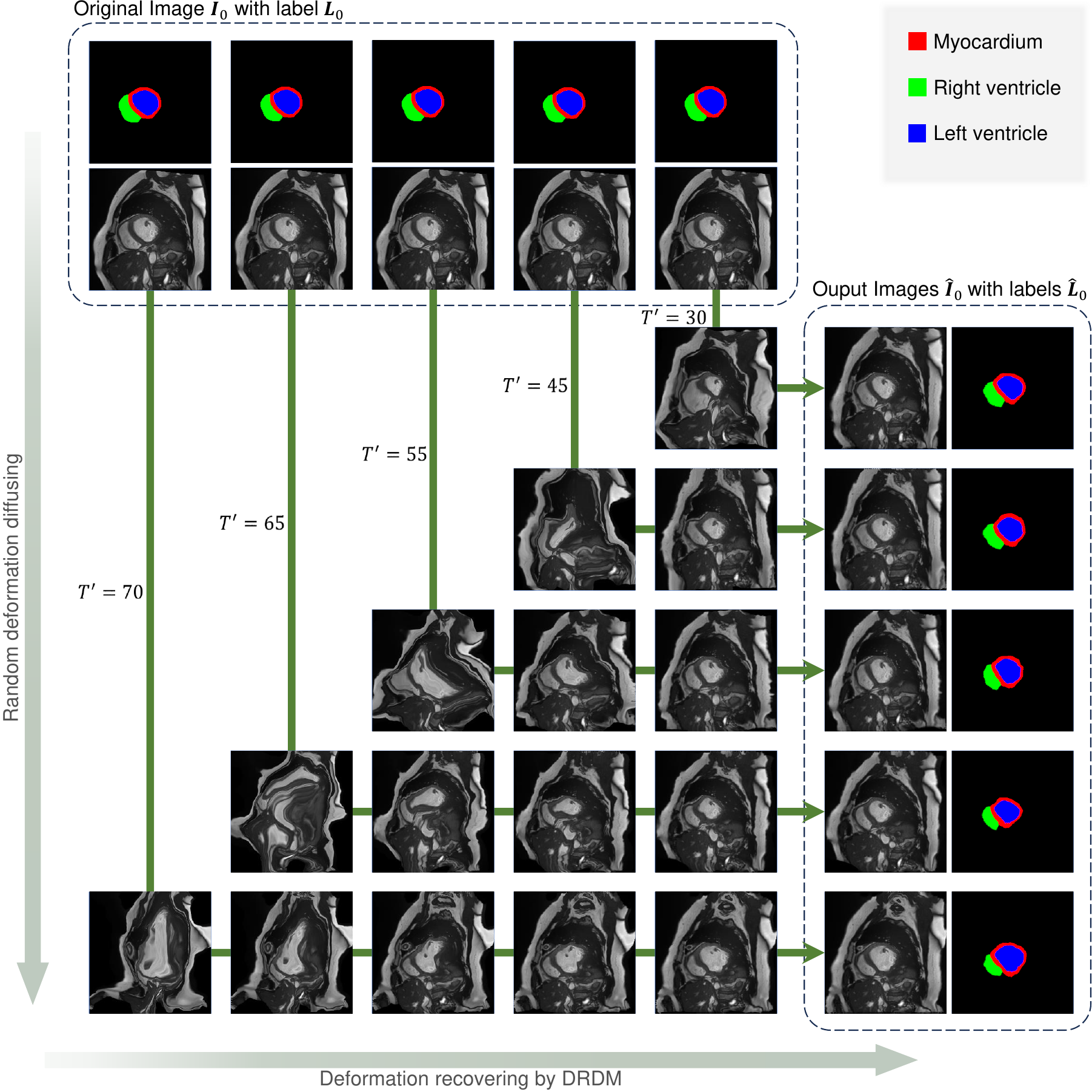}
\end{center}
\vspace{-.5em}
\caption{Visualisation of \textcolor{edited}{the} deformation diffusion and recovery \textcolor{edited}{processes} for 2D cardiac MRI images \textcolor{edited}{using the proposed} DRDM with varying \textcolor{edited}{deformation step levels} $T'$.}
\label{fig:data_diff_lvl}
\end{figure*}

\begin{figure*}[!h]
\begin{center}
\includegraphics[width=1.\linewidth]{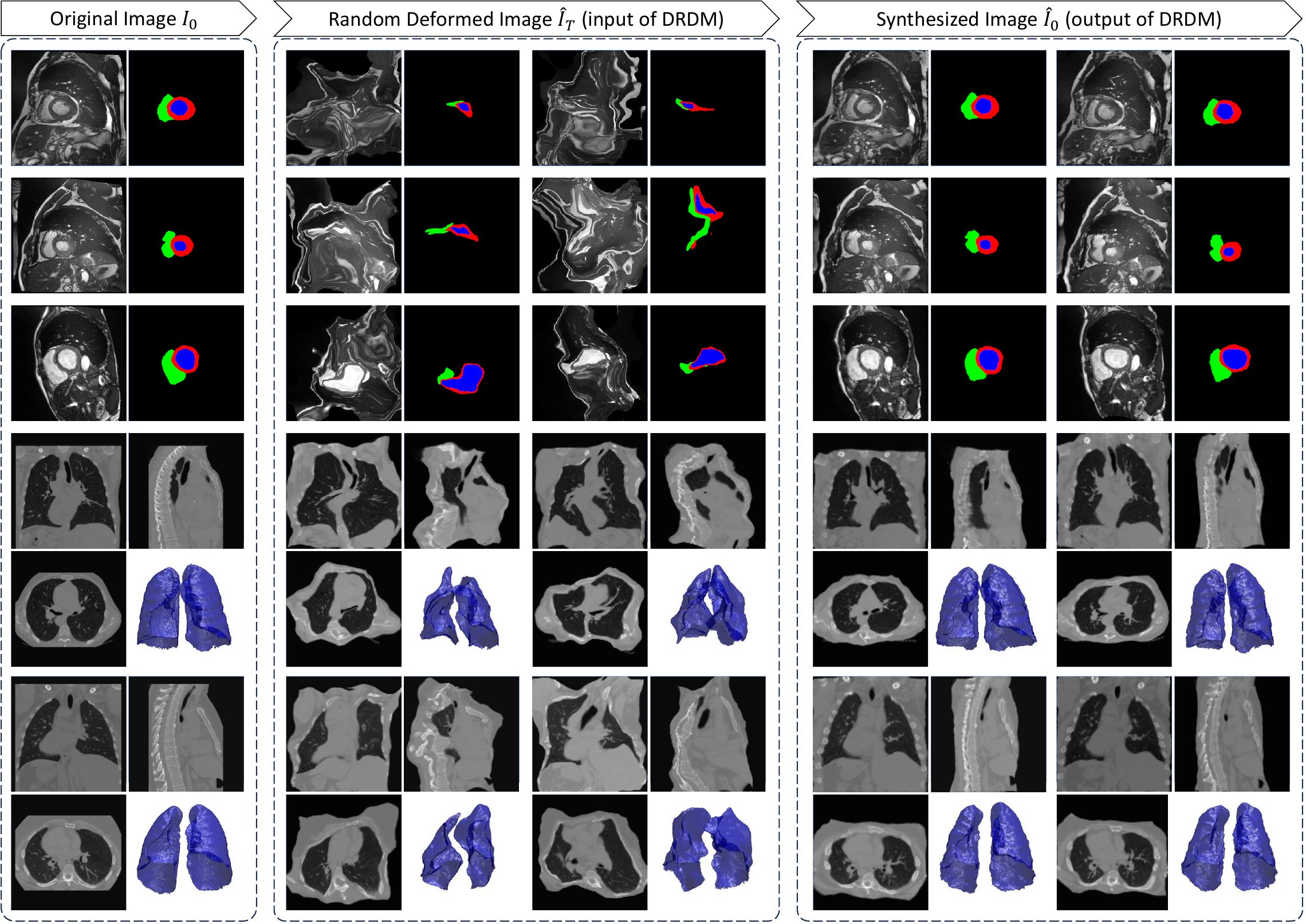}
\end{center}
\vspace{-.5em}
\caption{Diverse image deformation for 2D cardiac MRI and 3D pulmonary CT images (\textcolor{edited}{shown as} cross-sections through the image center in three orthogonal planes) \textcolor{edited}{using the proposed} DRDM. Left: original images, middle: \textcolor{edited}{randomly deformed images as inputs} of DRDM, and right: \textcolor{edited}{synthesized output images} from DRDM.}
\label{fig:data_diff_div}
\end{figure*}

\begin{figure*}[!th]
\begin{center}
\includegraphics[width=.85\linewidth]{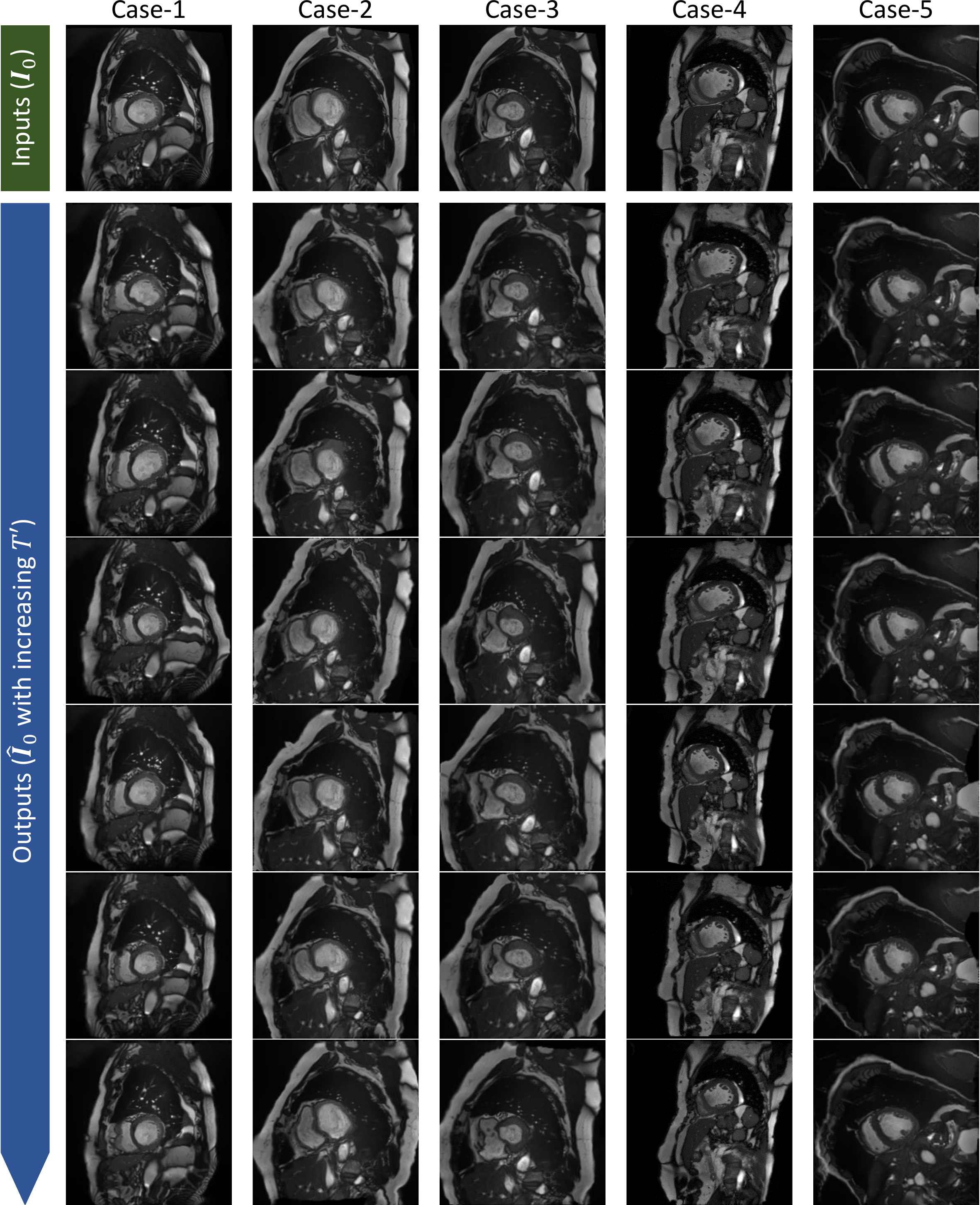}
\end{center}
\vspace{-.5em}
\caption{\textcolor{edited}{Original} and \textcolor{edited}{diversely} deformed images of five subjects \textcolor{edited}{using the proposed} DRDM for 2D cardiac MRI scans.}
\label{fig:examples_2d}
\end{figure*}

\begin{figure*}[!th]
\begin{center}
\includegraphics[width=.98\linewidth]{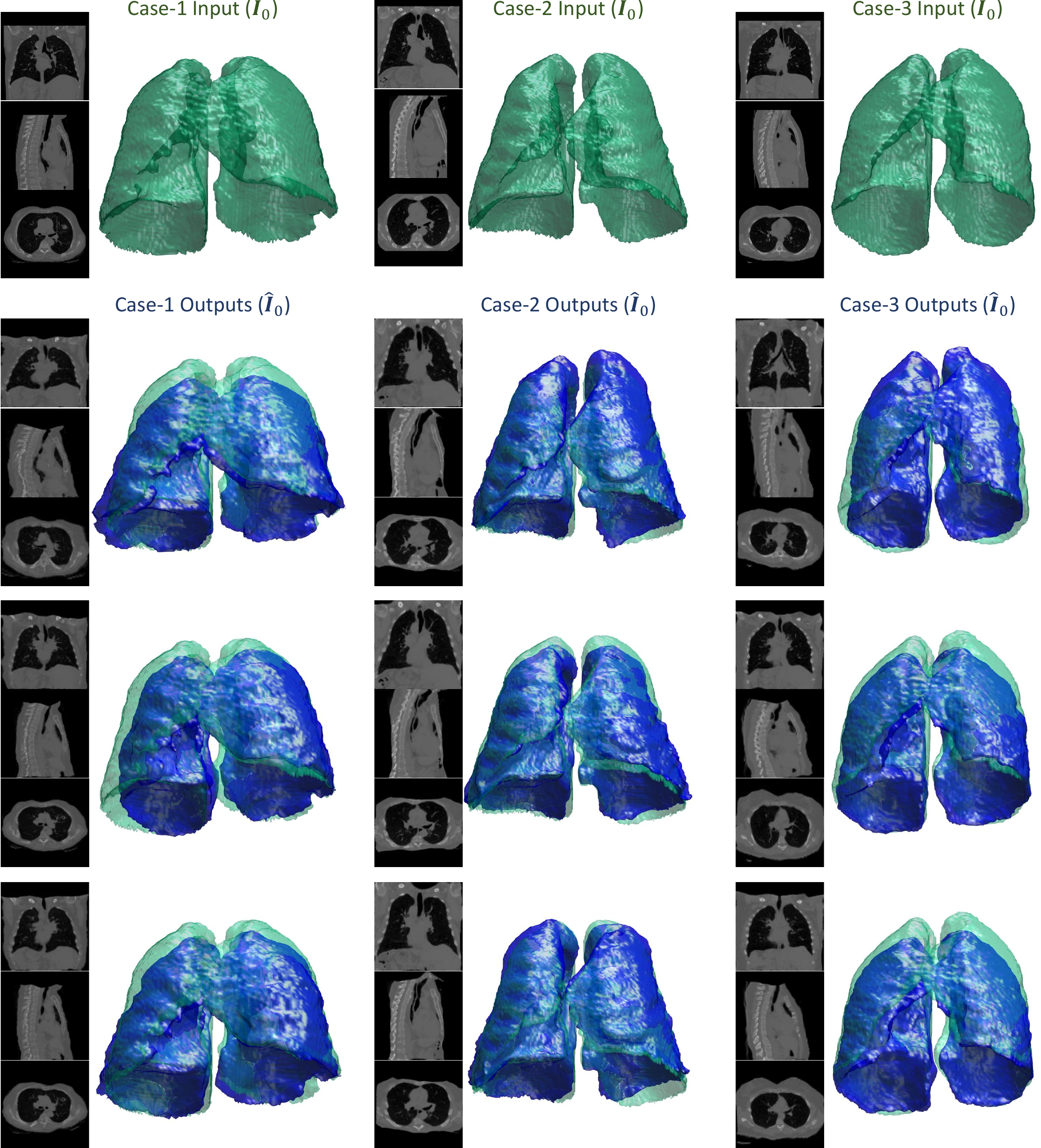}
\end{center}
\vspace{-.5em}
\caption{Lung shapes and \textcolor{edited}{three orthogonal cross-sections} (frontal, sagittal, and transverse) through the center of the image for original and deformed images (\textcolor[rgb]{0,0.7,0.4}{\rule{0.6em}{0.6em}} and \textcolor[rgb]{0,0.,0.7}{\rule{0.6em}{0.6em}}) of three subjects \textcolor{edited}{generated using the proposed} DRDM for 3D pulmonary CT scans.}
\label{fig:examples_3d}
\end{figure*}


\begin{table}[!htbp]
\caption{Average $\pm$ standard deviation \textcolor{edited}{of the magnitude} and the negative determinant ratio of Jacobian ($|{\textbf{\textit{J}}}|_{<0}$) \textcolor{edited}{for deformation fields genearted at} varying deformation level $T'$ in 2D cardiac MRI.}
\label{tab:def_mag_detj}
\centering 
\begin{tabular}{cccc}
\toprule
$T'$&$\max_x{{\|\phi[x]\|}_2}$&${\rm avg}_x{{\|\phi[x]\|}_2}$& ${|{\textbf{\textit{J}}}|_{<0}}$\\
(-)&(\%img~size)&(\%img~size)&(\textperthousand)\\
\midrule\midrule
30&7.8$\pm$1.3&2.4$\pm$0.6&0.7$\pm$1.5\\
45&10.4$\pm$2.4&3.0$\pm$0.8&0.9$\pm$1.2\\
55&11.8$\pm$3.2&3.3$\pm$0.8&2.4$\pm$3.2\\
65&12.7$\pm$4.0&3.6$\pm$0.9&6.6$\pm$5.5\\
70&14.8$\pm$3.5&4.3$\pm$1.4&9.8$\pm$7.2\\
\bottomrule
\end{tabular}
\end{table}

\section{Experiment of Image Deformation using {DRDM}}
\label{sec:img_syn}


As shown in Figure~\ref{fig:data_aug}(a), \textcolor{edited}{a small number of} 2D or 3D images are \textcolor{edited}{provided as inputs to the} \ac{DRDM} framework. 
The framework then generates \textcolor{edited}{corresponding} deformed images, with or without labels, for downstream tasks as described in the following Section~\ref{sec:aug_seg} and Section~\ref{sec:syn_reg}. 
\textcolor{edited}{For evaluation,} the datasets are divided into a source domain and a target domain. 
The diffusion networks of \ac{DRDM} are trained \textcolor{edited}{using} the source domain \textcolor{edited}{data} and then tested \textcolor{edited}{on} the target-domain \textcolor{edited}{data} for downstream tasks.
The datasets used in the experimental implementation of \ac{DRDM} are described in Section~\ref{sec:data}, data \textcolor{edited}{preprocessing methods are} explained in Section~\ref{sec:preproc}, \textcolor{edited}{the experimental setup is} detailed in Section~\ref{sec:set_drdm}, and the \textcolor{edited}{evaluation of generated data in presented} in Section~\ref{sec:results_img_synth}.

\subsection{Datasets}
\label{sec:data}


\textcolor{edited}{
To evaluate the effectiveness of the proposed method, two imaging modalities were used: \textit{cardiac MRI} and \textit{thoracic CT}. The \ac{DRDM} framework was trained independently on each modality and subsequently evaluated on both to assess its deformation generation and generalization performance.
}

\subsubsection{Cardiac MRI}
\textcolor{edited}{
Four public cardiac \ac{MRI} datasets were used to construct the training set (source domain), and one additional dataset was employed for downstream evaluation (target domain) to assess the performance of the proposed \ac{DRDM} framework. These datasets are as follows:}
\begin{enumerate}[a.]
\item The Sunnybrook Cardiac Data (SCD) \citep{radau2009evaluation} comprises 45 cine-\ac{MRI} images, representing a mix of \textcolor{edited}{subjects} with various conditions \textcolor{edited}{including} healthy individuals \textcolor{edited}{and patients} with hypertrophy, heart failure with \textcolor{edited}{and} without infarction.
\item  Task-6 of the Medical Segmentation Decathlon released as part of the Left Atrial Segmentation Challenge (LASC) \newline \cite{tobon2015benchmark}. 
It includes 30 3D \ac{MRI} volumes. 
\item  The Multi-Centre, Multi-Vendor \& Multi-Disease Cardiac Image Segmentation Challenge (M\&Ms dataset) \citep{campello2021multi} \textcolor{edited}{with} 375 patients with hypertrophic and dilated cardiomyopathies, as well as healthy subjects.
\item  Multi-Disease, Multi-View \& Multi-Center Right Ventricular Segmentation in Cardiac MRI (M\&Ms-2 dataset) \citep{martin2023deep} \textcolor{edited}{with} 360 patients with various right ventricle and left ventricle pathologies, as well as healthy subjects.
\item Automated Cardiac Diagnosis Challenge (ACDC) dataset \citep{bernard2018deep}, \textcolor{edited}{used for downstream evaluation in the whole-heart segmentation task, consists of 200 cardiac \ac{MRI} cases, with 100 for training and 100 for testing}.
\end{enumerate}
The datasets \textcolor{edited}{(a)-(d)} are used as source-domain data for training \ac{DRDM}, \textcolor{edited}{while} the dataset (e) \textcolor{edited}{served} as the target-domain data for downstream validation in the segmentation task as described in Section~\ref{sec:aug_seg}.

\subsubsection{Thoracic CT}
Following a similar approach to Cardiac MRI,  two \textcolor{edited}{publicly available} Thoracic CT datasets from the \textit{Cancer Imaging Archive}, \textcolor{edited}{were used to construct the training set (source domain), along with one additional dataset for downstream evaluation (target domain). These datasets are described as follows}:
\begin{enumerate}[a.]
\item  NSCLC-Radiomics (Version 4) \cite{aerts2015data}, which includes 422 \textcolor{edited}{patients diagnosed with} non-small cell lung cancer (NSCLC). 
\item QIN LUNG CT (Version 2) \cite{kalpathy2015qin}, which consists of 47 patients diagnosed with NSCLC at various stages and histologies.
\item The pulmonary \ac{CT} scans were provided by \cite{hering2022learn2reg} as part of the Learn2Reg 2021 challenge (task 2) dataset. 
These scans were consistently acquired at the same point within the breathing cycle to ensure uniformity. 
This dataset includes the inter-subject (exhale) registration task with 20 subjects for the training of a registration model and 10 for testing. Ground truth lung segmentations are also available for all scans. 
\end{enumerate}
The dataset (a) and (b) are used as source-domain data for training \ac{DRDM}, and the dataset c is used as the target-domain data for downstream validation in the inter-subject registration task as described in Section~\ref{sec:syn_reg}.

\subsection{Preprocessing and postprocessing}
\label{sec:preproc}

All images are first resized and padded to align with the isotropic resolution size of $ H \times W \times D $. The images are \textcolor{edited}{then} thresholded to remove \textcolor{edited}{irrelevant or non-anatomical} regions, such as \textcolor{edited}{air} cavity areas. \textcolor{edited}{Subsequently, image intensities were linearly normalized to the range [0, 1].}

To enhance the robustness of the \ac{DRDM} network, the images undergo several augmentations: rotation by a random angle $\sim\mathcal{U}(0^\circ, 180^\circ)$ around an arbitrary axis, translation by a random distance $\sim\mathcal{U}(-1/8, 1/8)$ of the image size along each of the three dimensions, randomly flipping with a probability of 0.5, and cropping with a ratio \textcolor{edited}{sampled from} $\sim\mathcal{U}(0.6, 1.0)$.

After deformation fields $\phi$ \textcolor{edited}{were} generated by \ac{DRDM}, they \textcolor{edited}{were} resized to the required \textcolor{edited}{resolution} $\tilde{\phi} \in \mathbb{R}^{H' \times W' \times D'}$, \textcolor{edited}{ensuring compatibility with the image and label dimensions required for downstream tasks}.

\subsection{Experimental setting for DRDM}
\label{sec:set_drdm}
\textcolor{edited}{During the} implementation of the random deformation diffusing process, \textcolor{edited}{thevectors in \acp{DDF} extending beyond} the field boundary could \textcolor{edited}{introduce numerical friction that limits} the increase in deformation magnitude. This occurs because the \textcolor{edited}{vectors outside the sampled region stop accumulating values when using} the "zero" padding mode.
\textcolor{edited}{To mitigate this issue, a larger deformation field was defined such that} $h>H,~w>W,~d>D$, 
and then the desired deformation field \textcolor{edited}{was cropped from} the centered region $H \times W \times D$ \textcolor{edited}{within} the created deformation field.

\textcolor{edited}{For all experiments}, $H, W, D$ and $H', W', D'$ are set the same \textcolor{edited}{values}, 256 for 2D \ac{MRI} scans and 128 for 3D \ac{CT} scans. The parameters $h, w, d$ are set to twice the dimensions of $H, W, D$, and $T$ is set at 80. To \textcolor{edited}{improve the robustness of \ac{DRDM}, a small amount of random noise was added during training, introducing a 5\% perturbation to the generated \acp{DVF}}. 

The noise level \textcolor{edited}{at} each time step is set as $n_{t} := \lfloor t^{0.6} \rfloor$ with $\alpha := 0.6$ and $\beta := 1$. As described in Section~\ref{sec:noisy_def_comp}, the theoretical setting \textcolor{edited}{for} the noise level for \textcolor{edited}{$\phi_{t:1}$} should be $n_{t:1} = \lfloor t^{1.6}/1.6 \rfloor$, \textcolor{edited}{however in practice, $n_{t:1}$ was set to $\lfloor t^{1.3}/1.5 \rfloor$ to mitigate the rounding effects and to} increase the redundancy range of network's prediction capacity, thus enhancing its ability to recover from random deformation at each step.

The training process uses \textcolor{edited}{the Adam optimizer with} $\lambda_1=1$, $\lambda_2=1$, and $\lambda_3=10$. \textcolor{edited}{The} initial learning rate of 0.0001 \textcolor{edited}{with} batch sizes of 64 for 2D and 4 for 3D. The \textcolor{edited}{model was trained for 1000 epochs on the 2D dataset and 2000 on} the 3D dataset to ensure the convergence. \textcolor{edited}{All experiments were performed on} An Intel Xeon(R) Silver 4210R CPU @ 2.40 GHz \ac{CPU} and an Nvidia Quadro RTX 8000 \ac{GPU} with 48 GB of memory.

\subsection{Image and deformation synthesis results}
\label{sec:results_img_synth}

\textcolor{edited}{An example of the deformation diffusion and recovery process for cardiac \ac{MRI} is shown in Figure~\ref{fig:data_diff_lvl}}. It shows that the deformation \textcolor{edited}{magnitude} becomes larger with increasing deformation level $T'$.
\textcolor{edited}{Additional examples of image synthesis for both cardiac \ac{MRI} and pulmonary \ac{CT} are presented in Figure~\ref{fig:data_diff_div}, illustrating that diverse images can be generated from a limited number of input \ac{MRI} and \ac{CT} images in both 2D and 3D}.

\textcolor{edited}{Further qualitative results are shown in Figure~\ref{fig:examples_2d} for 2D cardiac \ac{MRI} and Figure~\ref{fig:examples_3d} for 3D pulmonary \ac{CT} scans. These examples highlight the diversity and anatomical plausibility of the synthesized images. Comparisons with baseline methods are provided in Section~\ref{sec:def_base}, including BigAug~\citep{BigAug} (Figure~\ref{fig:examples_2d_base}) and the synthetic training method by Eppenhof and Pluim~\citep{eppenhof2018pulmonary} (Figure~\ref{fig:examples_3d_base})}.

Quantitative evaluation of the generated deformation is \textcolor{edited}{summarized} in Table~\ref{tab:def_mag_detj}. The \textcolor{edited}{maximum and average magnitudes of the deformation fields were measured as percentage of the image size and the ratio of voxels with a} negative Jacobian determinant of the deformation is used to \textcolor{edited}{assess deformation validity} of the generated DDF. The results in Table~\ref{tab:def_mag_detj} show the ratio of the negative Jacobian determinant ($\mathbb{E}_x{{\rm detJ}_{<0}(\phi)}$) and the magnitude ($\max_x{{\|\phi[x]\|}_2}$ and $\mathbb{E}_x{{\|\phi[x]\|}_2}$) of the generated deformation fields both increase with the larger deformation level $T'$. \textcolor{edited}{Nevertheless, the overall deformation quality remains} high ($\mathbb{E}_x{{\rm detJ}_{<0}(\phi)}<1\%$) even with a large deformation magnitude ($\max_x{{\|\phi[x]\|}_2}>10\%\times{H,W}$).

\textcolor{edited}{
These results confirm that the proposed \ac{DRDM} framework can generate large, diverse, and anatomically plausible diffeomorphic deformations while maintaining image quality. The quantitative metrics also provide practical guidance for selecting appropriate deformation magnitudes in different application scenarios.
}

\begin{algorithm}[!ht]
    \label{algo:data_aug}
    \caption{Data augmentation via \ac{DRDM}}
    \KwIn{Images and labels for deformation $\mathbf{D}^{\rm tgt}\subset{\mathbb{R}^{H\times W\times D}\times\mathbb{R}^{H\times W\times D\times C}}$}
    \KwOut{Deformed pairs of image and label $\mathbf{D}^{\rm aug}\subset{\mathbb{R}^{H\times W\times D}\times\mathbb{R}^{H\times W\times D\times C}}$}
    \vspace{-0.5em}
    \hrulefill\\
    \nl import the DRDM parameters $\theta$ from Algorithm~\ref{algo:train_drdm}\;
    \nl set a set of deformation levels $\mathcal{T}\subset\mathbb{Z}_+\cap[1,T]$\;
    \nl initialise the output set $\mathbf{D}^{\rm aug}\leftarrow\emptyset$\;
    \tcp{sample a pair of image and label}
    \nl \ForEach{$(\textbf{\textit{I}}_0,\textbf{\textit{L}}_0)\in\mathbf{D}^{\rm tgt}$}{
        \tcp{sample a deformation level number}
        \nl \ForEach{$T'\in\mathcal{T}$}{
        \nl generate \ac{DDF} $\phi$ using Algorithm~\ref{algo:infer_drdm}\;
        \nl deform the sampled image: $\hat{\textbf{\textit{I}}}\leftarrow \phi({\textbf{\textit{I}}}_0)$\;
        \nl deform the sampled label: $\hat{\textbf{\textit{L}}}\leftarrow \phi({\textbf{\textit{L}}}_0)$\;
        \nl append the deformed image and label into the output set: $\mathbf{D}^{\rm aug}\leftarrow \mathbf{D}^{\rm aug} \cup \{(\hat{\textbf{\textit{I}}},\hat{\textbf{\textit{L}}})\}$\;
        }
    }
    \nl return the output set $\mathbf{D}^{\rm aug}$.
\end{algorithm}

\section{Downstream application in image segmentation}
\label{sec:aug_seg}

\textcolor{edited}{As illustrated in} Figure~\ref{fig:data_aug}(b), the generated images \textcolor{edited}{and their} corresponding labels can be used for training a segmentation model $\mathcal{S}$. \textcolor{edited}{In this section, \ac{DRDM} is evaluated as a data augmentation framework for few-shot learning in medical image segmentation.}
The segmentation framework is described in Section~\ref{sec:seg_framework}, with the training process described in Section~\ref{sec:seg_train}. The experimental setup and the corresponding results are explained in Section~\ref{sec:seg_exp_set} and Section~\ref{sec:seg_results}, respectively.

\subsection{Segmentation framework}
\label{sec:seg_framework}

In the segmentation framework, segmentation masks of specific regions
${\textbf{\textit{L}}}$ are \textcolor{edited}{predicted} by a segmentation network $\mathcal{S}_{\zeta}$ from a given image $\textbf{\textit{I}}$:
\begin{equation}
    \mathcal{S}_\zeta: \mathbb{R}^{H \times W \times D}\to\{0,1\}^{H \times W \times D\times C},~\textbf{\textit{I}} \mapsto \tilde{\textbf{\textit{L}}}
\end{equation}
with the trainable parameters $\zeta$ optimized by \textcolor{edited}{minimizing the following}:
\begin{equation}
\min_{\zeta}{\{\mathcal{L}_{\rm seg}({\textbf{\textit{L}}},\tilde{\textbf{\textit{L}}})\}}
\end{equation}
where $c$ denotes \textcolor{edited}{the number of output channels}, $\tilde{\textbf{\textit{L}}}$ denotes the \textcolor{edited}{predicted segmentation mask}.

The U-Net \citep{ronneberger2015u}, is \textcolor{edited}{employed} as the segmentation model in this experiment. \textcolor{edited}{Each} dense block consists of two 3x3 convolutions, each followed by a \ac{ReLU}. \textcolor{edited}{The encoder path includes max-pooling operations with a stride of 2, while the decoder path employs up-sampling operations with the same stride. A Sigmoid activation function is used at the network output to produce probabilistic segmentation masks.}
\subsection{Segmentation network \textcolor{edited}{training}}
\label{sec:seg_train}

The segmentation loss function $\mathcal{L}_{\rm seg}$ is based on \ac{BCE}:
\begin{equation}
\label{eq:loss_seg}
\left\{
\begin{array}{l}
\mathcal{L}_{\rm seg}:=\mathbb{E}_{x}\left(\sum_{i=1}^{C}{\mathcal{L}_{\rm seg}^{\rm bce}(\textbf{\textit{L}}[x,i],\tilde{\textbf{\textit{L}}}[x,i])}\right)\\
\mathcal{L}_{\rm seg}^{\rm bce}:=\textbf{\textit{L}}[x,i]\log(\tilde{\textbf{\textit{L}}}[x,i])
+(1-\textbf{\textit{L}}[x,i])\log(1-\tilde{\textbf{\textit{L}}}[x,i])\\
\end{array}\right.
\end{equation}

\textcolor{edited}{The network is trained using the Adam optimizer with an initial learning rate of 0.001 and an exponential learning rate scheduling strategy. The batch size is set to 64 for 2D images. Training is conducted using the same computational hardware described in Section~\ref{sec:set_drdm}.}

\subsection{Experimental setup for segmentation}
\label{sec:seg_exp_set}

\textcolor{edited}{Within the segmentation framework, the original images and corresponding labels from the target domain were augmented using \ac{DRDM} with varying deformation levels $T'$, as illustrated in Figure~\ref{fig:data_aug}(b). The augmented data were included in the training set for downstream segmentation tasks.}

\textcolor{edited}{To assess the effectiveness of \ac{DRDM} for data augmentation, several augmentation strategies were compared within the same segmentation framework.}
\textcolor{edited}{Among them, BigAug~\citep{BigAug} was selected as the primary baseline. BigAug applies nine stacked transformation modules that alter image quality, appearance, and spatial characteristics (including deformation) to improve domain generalization performance.}

As described in Section~\ref{sec:data}, the ACDC data \textcolor{edited}{divided into 100 subjects for training and 100 for testing}. 
\textcolor{edited}{In the segmentation experiments, each dataset was augmented 32 times using both \ac{DRDM}, following Algorithm~\ref{algo:data_aug}, and BigAug~\citep{BigAug}.}
\textcolor{edited}{Comparisons were performed under different training conditions, with labeled data subsets containing 5 subjects (5\%), 20 subjects (20\%), 50 subjects (50\%), and 100 subjects (100\%).}

\textcolor{edited}{Segmentation models trained with different augmentation strategies were evaluated using multiple performance metrics, including the \ac{ASD}, \ac{DSC} (F1-score), precision (F0-score), and sensitivity (F$\infty$-score) between the ground-truth segmentation masks $\textbf{\textit{L}}$ and the predicted results $\tilde{\textbf{\textit{L}}}$.}

Ground-truth label \textcolor{edited}{maps} were created where 0, 1, 2 and 3 represent voxels located in the background, in the RV cavity, in the myocardium, and in the LV cavity.

\begin{figure}[!thb]
\begin{center}
\includegraphics[width=\linewidth]{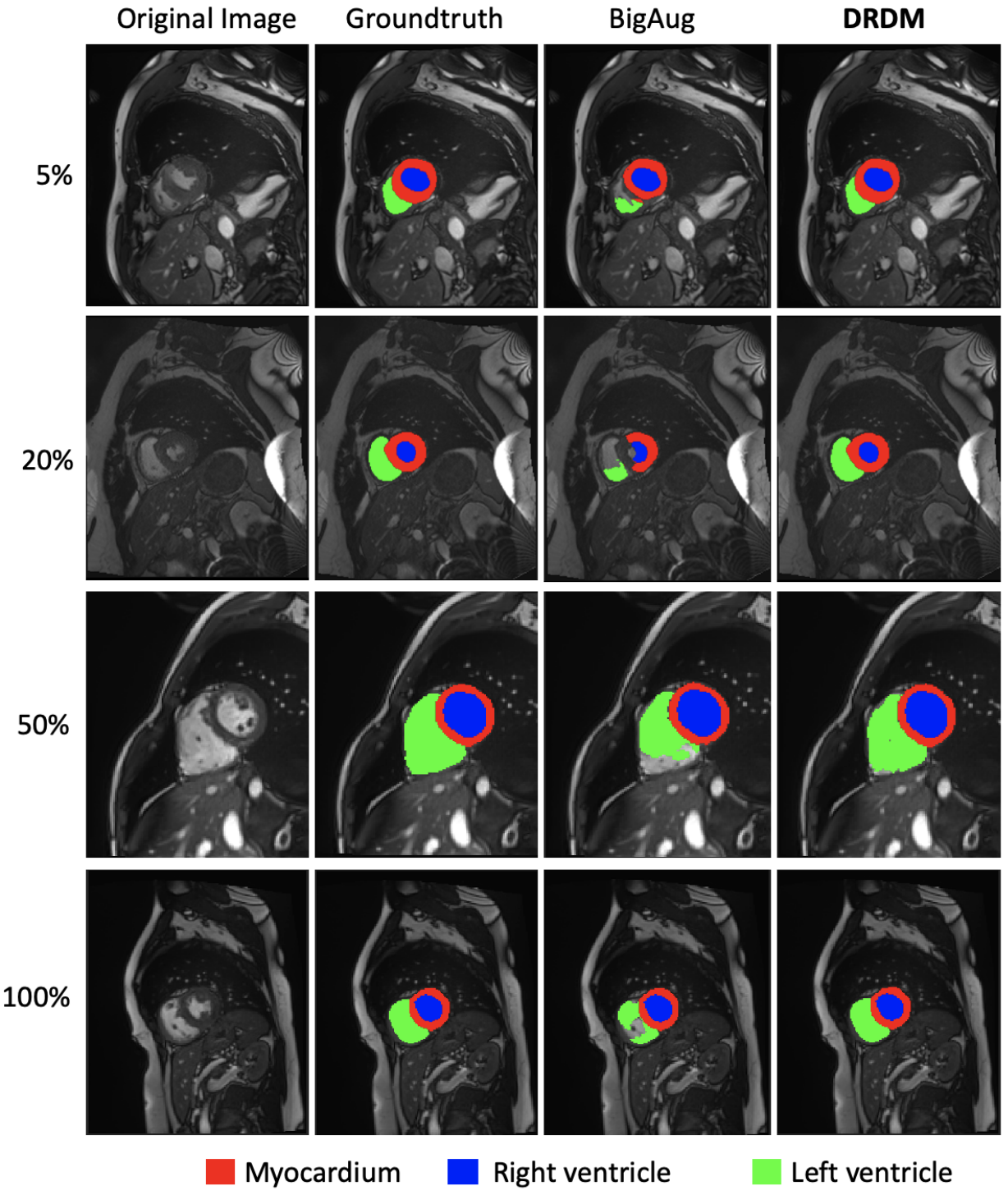}
\end{center}
\vspace{-.5em}
\caption{Segmentation \textcolor{edited}{results of models trained with varying ratios} of labeled data, comparing different augmentation methods based on BigAug and DRDM. \textcolor{edited}{updated as required by Comment-2.5}}
\label{fig:seg_examp}
\end{figure}

\begin{figure}[!thb]
\begin{center}
\includegraphics[width=\linewidth]{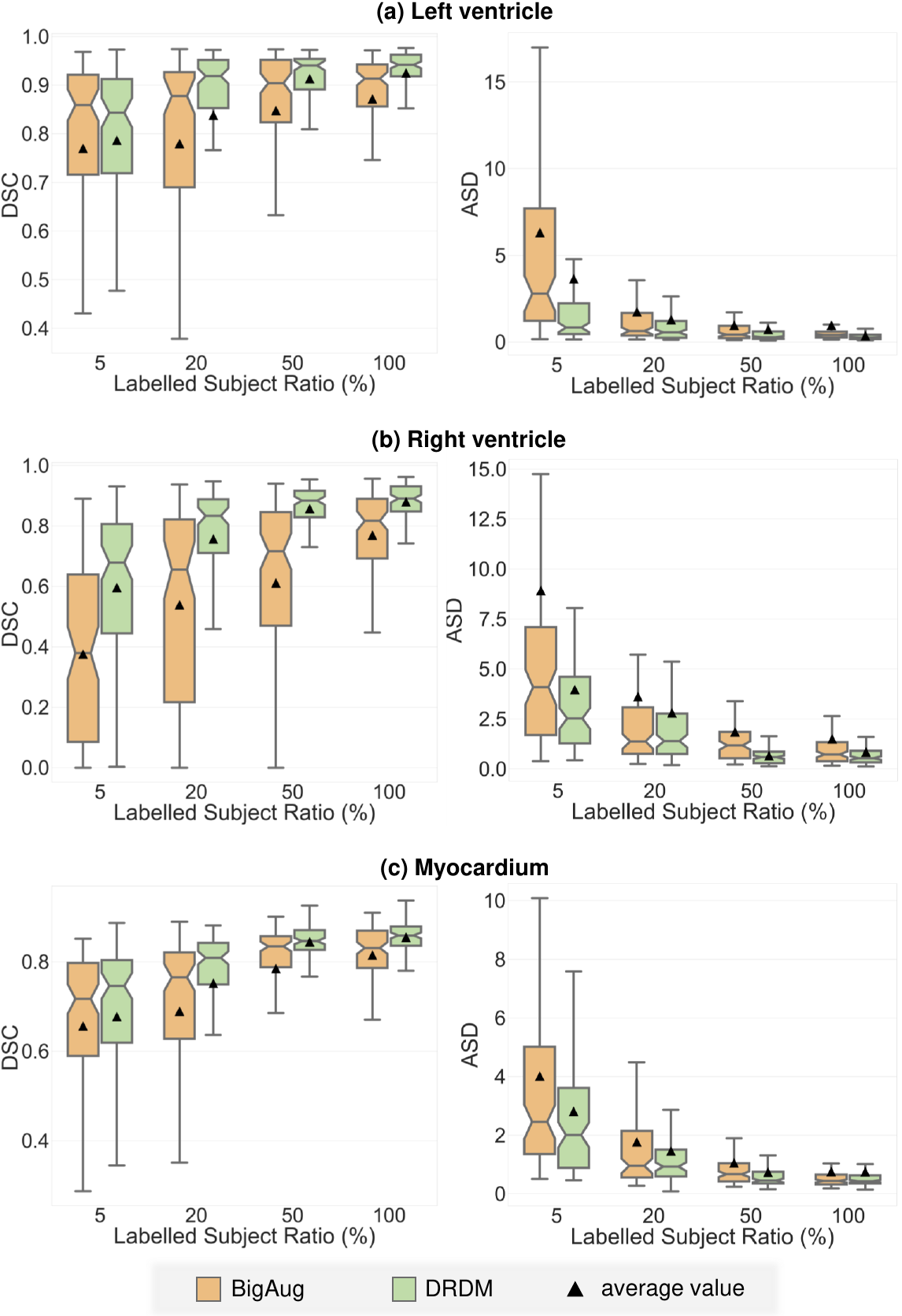}
\end{center}
\vspace{-.5em}
\caption{Quantitative \textcolor{edited}{segmentation results of models trained with the varying ratios of labelled data, comparing the proposed DRDM framework with baseline methods across three cardiac structures in MRI. The results demonstrate that {DRDM} consistently} outperforms the baseline under different levels of labelled data \textcolor{edited}{availability}.}
\label{fig:seg_box_plot}
\end{figure}

\begin{table*}[!thb]
\caption{Segmentation results \textcolor{edited}{on cardiac MRI showing the} average DSC (\%), sensitivity (sns/\%), and precision (prec/\%) \textcolor{edited}{obtained using} different data augmentation methods. \textcolor{edited}{Results are reported for a vanilla U-Net trained} with varying subjects number (\#subj) and ratio of the labelled images.}
\label{tab:seg_results_2d}
\centering 
\begin{tabular}{cccccccccccccc}
\toprule
\multirow{2}{*}{\cellcolor{white} \makecell{\#subj\\/ratio}}&\multirow{2}{*}{\makecell{aug\\method}}&\multicolumn{3}{c}{Left ventricle}&\multicolumn{3}{c}{Right ventricle}&\multicolumn{3}{c}{Myocardium}&\multicolumn{3}{c}{Average}\\
\cmidrule(lr){3-5}  \cmidrule(lr){6-8} \cmidrule(lr){9-11} \cmidrule(lr){12-14}
&& {dsc$\uparrow$}&{sns$\uparrow$}&{prec$\uparrow$}
& {dsc$\uparrow$}&{sns$\uparrow$}&{prec$\uparrow$}
& {dsc$\uparrow$}&{sns$\uparrow$}&{prec$\uparrow$}
& dsc$\uparrow$&sns$\uparrow$&prec$\uparrow$\\
\midrule
\midrule
& N/A &   55.0&\textbf{77.5}&45.5&39.5&48.9&35.6 &46.3&63.9&39.0&46.9&63.4&40.0\\
& BigAug &75.8&73.6&{83.2}&       37.6&29.6 &65.2&65.6&66.0&69.1&59.7&57.7&72.5\\
\rowcolor[RGB]{230,230,230} \cellcolor{white!0}
\multirow{-3}{*}{\makecell{\cellcolor{white!0} 5 /\\ \cellcolor{white!0} 5\%}} 
& DRDM & \textbf{77.0}&{73.7} &\textbf{84.1}& \textbf{59.6}&\textbf{55.8}&\textbf{77.3}&\textbf{67.9}&\textbf{68.5}&\textbf{73.4}&\textbf{68.2}&\textbf{66.0}&\textbf{78.3}\\
\midrule
& N/A &75.8&83.7&72.8&59.7& 52.6& 74.9&62.1& 63.7&66.8&65.8&66.6& 71.5\\
& BigAug & 78.0&75.3&\textbf{85.5}& 53.9&47.1 &\textbf{79.9}&68.9&64.1&\textbf{81.6}&66.9&62.2&\textbf{82.3}\\
\rowcolor[RGB]{230,230,230} \cellcolor{white!0}
\multirow{-3}{*}{\makecell{\cellcolor{white!0} 20 /\\ \cellcolor{white!0} 20\%}} 
& DRDM & \textbf{83.0}&\textbf{86.8} &81.4&\textbf{75.0}&\textbf{75.6}&78.9&\textbf{74.6}&\textbf{82.5}&71.8&\textbf{77.5}&\textbf{82.6}&77.4\\
\midrule
& N/A &  82.9&80.6 &83.0& 70.2&65.1&83.3&74.6&77.9 &72.7&75.9&74.5&79.7\\
& BigAug & 84.8&83.1&\textbf{90.3}&  61.1&54.5 &\textbf{84.9}&78.5&77.1&82.8&74.8&71.6&86.0\\
\rowcolor[RGB]{230,230,230} \cellcolor{white!0}
\multirow{-3}{*}{\makecell{\cellcolor{white!0} 50 /\\ \cellcolor{white!0} 50\%}} 
& DRDM &  \textbf{91.4}&\textbf{95.5} &{88.3}&   \textbf{85.6}&\textbf{89.8} &{83.5}&\textbf{84.1}&\textbf{94.4} &\textbf{88.3}&\textbf{87.1}&\textbf{93.2}&\textbf{86.7}\\
\midrule
& N/A &   89.3&90.5 &{92.6} &85.2&83.2&{89.4}&84.9&83.8&\textbf{86.9}&86.5& 85.8&89.6 \\
& BigAug &87.2&80.2&\textbf{97.6}& 76.9& 69.3&\textbf{92.7}&81.5&79.8&{85.0}&81.9&76.4&\textbf{91.8}\\
\rowcolor[RGB]{230,230,230} \cellcolor{white!0}
\multirow{-3}{*}{\makecell{\cellcolor{white!0} 100 /\\ \cellcolor{white!0} 100\%}} 
& DRDM & \textbf{92.5} & \textbf{96.5}&{89.1}&   \textbf{87.9} & \textbf{93.2}&{83.8}&\textbf{85.4} & \textbf{95.1}&{77.8}&\textbf{88.6}&\textbf{94.9}&83.6\\
\bottomrule
\end{tabular}
\end{table*}

\subsection{Image segmentation results}
\label{sec:seg_results}

\textcolor{edited}{Representative segmentation results on} cardiac \ac{MRI} are shown in Figure~\ref{fig:seg_examp}, \textcolor{edited}{for different} ratios of labelled data at 5\%, 20\%, 50\%, and 100\%. These qualitative results demonstrate that the U-Net model augmented with \textcolor{edited}{the proposed \ac{DRDM} framework} outperforms the BigAug approach, particularly in the \textcolor{edited}{segmentation of the} right ventricle.

The distribution of \ac{DSC} and \ac{ASD} values are presented in Figure~\ref{fig:seg_box_plot} for further quantitative comparison. The results indicate that \textcolor{edited}{the proposed} DRDM method outperforms BigAug across most label ratio settings. Specifically, the \ac{DSC} and \ac{ASD} metrics for \textcolor{edited}{the proposed} \ac{DRDM} are significantly higher ($p<0.01$) than those for BigAug in the right ventricle and generally \textcolor{edited}{improved} in the other \textcolor{edited}{cardiac structures}.

\textcolor{edited}{Quantitative results} for \ac{DSC}, sensitivity, and precision are presented in Table~\ref{tab:seg_results_2d}. These results consistently show that \ac{DRDM} outperforms BigAug in most settings for \ac{DSC} and \ac{sns}. \textcolor{edited}{Notably,} BigAug tends to conservatively segment \textcolor{edited}{cardiac structures}, as shown in Figure~\ref{fig:seg_examp}, resulting in higher precision but \textcolor{edited}{an increased false-negative rate, and consequently lower} sensitivity.


\begin{algorithm}[!ht]
    \label{algo:data_syn}
    \caption{Data synthesis via \ac{DRDM}}
    \KwIn{Images for deformation $\mathbf{D}^{\rm tgt}\subset{\mathbb{R}^{H\times W\times D}}$}
    \KwOut{Pairs of deformed images and the \ac{DDF} $\mathbf{D}^{\rm syn}\subset{\mathbb{R}^{H\times W\times D}\times\mathbb{R}^{H\times W\times D}\times\mathbb{R}^{H\times W\times D\times 3}}$}
    \vspace{-0.5em}
    \hrulefill\\
    \nl import the DRDM parameters $\theta$ from Algorithm~\ref{algo:train_drdm}\;
    \nl set a set of deformation levels $\mathcal{T}\subset\mathbb{Z}_+\times\mathbb{Z}_+$\;
    \nl initialise the output set $\mathbf{D}^{\rm syn}\leftarrow\emptyset$\;
    \tcp{sample an image}
    \nl \ForEach{$\textbf{\textit{I}}_0\in\mathbf{D}^{\rm tgt}$}{
        \tcp{sample deformation level numbers}
        \nl \ForEach{$(T'_{\rm aug},T'_{\rm syn})\in\mathcal{T}$}{
        \tcp{create the moving image}
        \nl generate \ac{DDF} $\phi_{\rm aug}$ based on $(\textbf{\textit{I}}_0,T'_{\rm aug})$ using Algorithm~\ref{algo:infer_drdm}\;
        \nl deform the sampled image: ${\textbf{\textit{I}}}^{\rm mv}\leftarrow \phi_{\rm aug}({\textbf{\textit{I}}}_0)$\;
        \tcp{create the fixed image and DDF}
        \nl generate \ac{DDF} $\phi_{\rm syn}$ based on $(\textbf{\textit{I}}^{\rm mv},T'_{\rm syn})$ using Algorithm~\ref{algo:infer_drdm}\;
        \nl deform the sampled image: ${\textbf{\textit{I}}}^{\rm fx}\leftarrow \langle\phi_{\rm syn}\circ\phi_{\rm aug}\rangle({\textbf{\textit{I}}}_0)$\;
        \nl append the deformed images and the DDF: $\mathbf{D}^{\rm aug}\leftarrow \mathbf{D}^{\rm syn} \cup \{({\textbf{\textit{I}}}^{\rm mv},{\textbf{\textit{I}}}^{\rm fx},\phi_{\rm syn})\}$\;
        }
    }
    \nl return the output set $\mathbf{D}^{\rm syn}$.
\end{algorithm}


\begin{figure}[!th]
\begin{center}
\includegraphics[width=\linewidth]{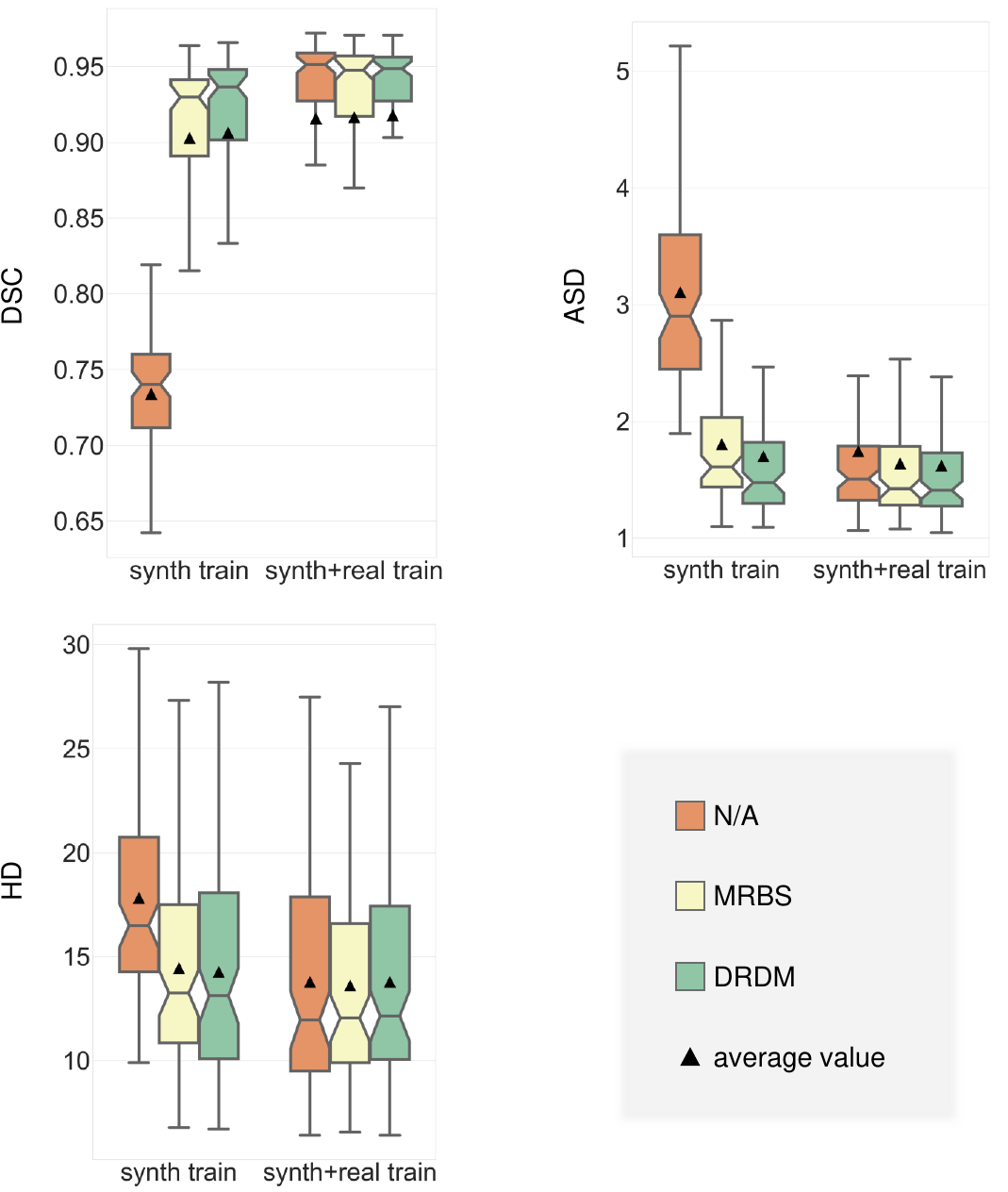}
\end{center}
\vspace{-.5em}
\caption{Quantitative results of registration models \textcolor{edited}{trained with synthetic and real data}, demonstrating the performance improvement achieved through pretraining using the proposed DRDM framework.}
\label{fig:reg_box_plot}
\end{figure}

\section{Downstream application in image registration}
\label{sec:syn_reg}

As illustrated in Figure~\ref{fig:data_aug}(b), the \textcolor{edited}{synthesised} images and their corresponding \acp{DDF} \textcolor{edited}{generated by \ac{DRDM} are used to pre-train} a registration model $\mathcal{R}$. \textcolor{edited}{In this section}, \ac{DRDM} is \textcolor{edited}{evaluated} as a data synthesis tool for synthetic training \textcolor{edited}{in} the image registration task. 
The \textcolor{edited}{registration} framework is described in Section~\ref{sec:reg_framework}, with the training process described in Section~\ref{sec:reg_train}. The experimental setup and the corresponding results are explained in Section~\ref{sec:reg_exp_set} and Section~\ref{sec:reg_results}, \textcolor{edited}{respectively}.

\subsection{Registration framework}
\label{sec:reg_framework}

In the registration framework, the deformation field between a pair of images is estimated by a registration network $\mathcal{R}_{\eta}$ \textcolor{edited}{defined as}:
\begin{equation}
\mathcal{R}_{\eta}: (\mathbb{R}^{H\times W\times D},\mathbb{R}^{H\times W\times D})\to \mathbb{R}^{H\times W\times D\times 3},~ (\textbf{\textit{I}}^{\rm mv},\textbf{\textit{I}}^{\rm fx})\mapsto \tilde{\phi}
\end{equation}
with the trainable parameters $\eta$ optimized by \textcolor{edited}{minimizing the following}:
\begin{equation}
\min_{\eta}{\{\mathcal{L}_{\rm reg}({\textbf{\textit{I}}}^{\rm mv},{\textbf{\textit{I}}}^{\rm fx},\tilde{\phi})\}}
\end{equation}
where ${\textbf{\textit{I}}}^{\rm mv},{\textbf{\textit{I}}}^{\rm fx}$ denotes the moving and the fixed images, \textcolor{edited}{respectively}, and $\tilde{\phi}$ denotes the estimated deformation field.

\textcolor{edited}{In this study, the VoxelMorph architecture~\citep{balakrishnan2019voxelmorph} is employed as the registration network, providing a robust and widely adopted baseline for learning-based image registration.}

\subsection{Registration network \textcolor{edited}{training}}
\label{sec:reg_train}
The registration network is \textcolor{edited}{pretrained using} synthetic image \textcolor{edited}{pairs and their} corresponding deformation \textcolor{edited}{fields}, which are \textcolor{edited}{generated} using \ac{DRDM}, with varying deformation levels $T'$.
The \textcolor{edited}{pretraining} registration loss $\mathcal{L}_{\rm reg}$ consists of two components, \ac{MSE} and \textcolor{edited}{a regularization} term:
\begin{equation}
\label{eq:loss_reg}
\left\{
\begin{array}{l}
\mathcal{L}_{\rm reg}:=\lambda_{4}\mathcal{L}_{\rm reg}^{\rm mse}+\lambda_{5}\mathcal{L}_{\rm reg}^{\rm grad}\\
\mathcal{L}_{\rm reg}^{\rm mse}:=\mathbb{E}_{x}{({\|\phi-\tilde{\phi}\|}_{2})}\\
\mathcal{L}_{\rm reg}^{\rm grad}:=\mathbb{E}_{x}({\|\nabla\tilde{\phi}[x]\|}_1)\\
\end{array}\right.
\end{equation}
\textcolor{edited}{Following pretraining, the registration model is fine-tuned using the optimization strategy proposed by~\citep{balakrishnan2019voxelmorph}}.

The synthetic training process uses $\lambda_4=1$ and $\lambda_5=1$, with the Adam optimizer. \textcolor{edited}{The} initial learning rate \textcolor{edited}{is set to} 0.0001 and the batch sizes of 12. \textcolor{edited}{Training is performed on the same computational hardware described in Section~\ref{sec:set_drdm}.}

\subsection{Experimental setup for registration}
\label{sec:reg_exp_set}

As described in Section~\ref{sec:data}, the pulmonary \ac{CT} data provided by \citep{hering2022learn2reg} is split into 20 for training and 10 for testing.

\textcolor{edited}{Following} Algorithm~\ref{algo:data_syn}, the original images are first augmented \textcolor{edited}{using} \ac{DRDM} \textcolor{edited}{to generate} moving images $\textbf{\textit{I}}^{\rm mv}$, and then \textcolor{edited}{further} deformed by \ac{DRDM} \textcolor{edited}{to produce} fixed images $\textbf{\textit{I}}^{\rm fx}$ with \textcolor{edited}{corresponding} deformation fields $\phi$. \textcolor{edited}{These image pairs, along with their associated deformation fields, were used as synthetic training data for the registration model.}

To \textcolor{edited}{evaluate the effectiveness} of \ac{DRDM} in image registration, \textcolor{edited}{the proposed framework was compared with the synthetic training approach of~\citep{eppenhof2018pulmonary}, which is based on a model-registered B-spline (\ac{MRBS}) deformation generation method. The same experimental configuration as described in~\citep{eppenhof2018pulmonary} was adopted for a fair comparison}.

\textcolor{edited}{In accordance with that setting, 20 \ac{CT} scans were each augmented 32 times, and the deformation fields to be learned were synthesized using either \ac{DRDM} or the B-spline transformer. Following synthetic training, all registration models were further fine-tuned in an unsupervised manner using the real pulmonary \ac{CT} data, following the optimization strategy of~\citep{balakrishnan2019voxelmorph}.}

\textcolor{edited}{Registration performance was evaluated using multiple quantitative metrics}, including \ac{DSC} (F1), \ac{ASD}, and \ac{HD} \textcolor{edited}{computed between the ground-truth lung masks in the fixed image $\textbf{\textit{L}}$ and the corresponding deformed masks $\tilde{\textbf{\textit{L}}}$ obtained via the estimated deformation field}.

\begin{table}[!thb]
\caption{Inter-subject registration results \textcolor{edited}{on pulmonary CT showing the} average DSC (\%), ASD (voxel), and HD (voxel) using a vanilla VoxelMorph \citep{balakrishnan2019voxelmorph}.
\textcolor{edited}{Results are reported for different synthetic training strategies, including the Multi-Resolution B-Spline (MRBS) method~\citep{eppenhof2018pulmonary} and the proposed \ac{DRDM} framework, followed by unsupervised fine-tuning on real training data}.}
\label{tab:reg_results_3d}
\centering 
\begin{tabular}{cccccc}
\toprule
\multirow{2}{*}{\makecell{synth\\method}}
& \multirow{2}{*}{\makecell{real\\train}}
& DSC$\uparrow$&ASD$\downarrow$&HD$\downarrow$&${|{\textbf{\textit{J}}}|_{<0}}\downarrow$\\
&&(\%)&(vox)&(vox)&(\textperthousand)\\
\midrule
\midrule
N/A&$\times$
& 73.39 &3.11&17.82&-\\
MRBS &$\times$& {90.29}  &{1.80}&{14.45}&\textbf{3.25}\\
\rowcolor[RGB]{230,230,230}
DRDM &$\times$& \textbf{90.64}  &\textbf{1.71}&\textbf{14.27}&{4.42}\\
\hline
N/A &\checkmark& 91.57  &1.74&13.80& 5.38\\
MRBS &\checkmark&91.66   &1.64&\textbf{13.62}&4.96\\
\rowcolor[RGB]{230,230,230}
DRDM &\checkmark& \textbf{91.79}  &\textbf{1.62}&13.71&\textbf{4.95}\\

\bottomrule
\end{tabular}
\end{table}

\subsection{Image registration results}
\label{sec:reg_results}

The distribution of \ac{DSC}, \ac{ASD}, and \ac{HD} values evaluated for \textcolor{edited}{the proposed} model and the baseline \textcolor{edited}{models} are \textcolor{edited}{shown} in Figure~\ref{fig:reg_box_plot}. The registration model synthetically trained by \textcolor{edited}{the proposed} \ac{DRDM} method outperforms \textcolor{edited}{the model trained with the \ac{MRBS}-based method} in \ac{ASD} ($p<0.05$). \textcolor{edited}{Furthermore, the model pretrained synthetically with \ac{DRDM} achieves performance comparable to that of the model trained directly on real data.}

\textcolor{edited}{Quantitative results, including the} average values of \ac{DSC}, \ac{ASD}, \ac{HD}, and negative Jacobian determinant ratio are \textcolor{edited}{summarised} in Table~\ref{tab:reg_results_3d}. These results consistently demonstrate that \textcolor{edited}{the proposed \ac{DRDM} framework significantly ($p<0.05$)} outperforms \ac{MRBS} in synthetic training of the registration model, achieving \textcolor{edited}{registration accuracy comparable to real-data training}. 

\textcolor{edited}{Overall, these findings further confirm the effectiveness of the deformed images generated by \ac{DRDM} for enhancing synthetic training in image registration tasks.}

\section{Related Works}
\label{sec:related_works}

\subsection{Diffusion models in medical image analysis}

\textcolor{edited}{Several recent studies have investigated the application of diffusion models to various} medical image analysis tasks, including anomaly detection \cite{wolleb2022diffusion,bercea2023mask,liang2023modality} and image registration \citep{qin2023fsdiffreg,gao2023diffusing}.

\cite{wolleb2022diffusion} proposed a method \textcolor{edited}{combining} an intensity noising-denoising scheme \citep{ho2020denoising,song2020denoising} with classifier guidance for 2D image-to-image translation. This technique transforms diseased \textcolor{edited}{subjects' images} into their healthy counterparts while preserving anatomical information, \textcolor{edited}{allowing the difference between the original and translated images to highlight anomaly regions in brain MRI}. 
Similarly, \cite{bercea2023mask} introduced an AutoDDPM method, based on \ac{DDPM} \citep{ho2020denoising}, for anomaly detection in brain MRI, incorporating an iterative process of stitching-and-resampling to generate pseudo-healthy images. 
\textcolor{edited}{In another study, \cite{liang2023modality} proposed MMCCD for multimodal brain MRI anomaly segmentation}, utilizing an intensity-based diffusion model \citep{ho2020denoising,song2020denoising}.

\cite{qin2023fsdiffreg} integrated \ac{DDPM} \citep{ho2020denoising} \textcolor{edited}{into} a registration framework, introducing \textcolor{edited}{two complementary diffusion modules}: feature-wise diffusion-guided module to enhance feature processing during the registration process, and a score-wise diffusion module to guide the optimisation process while preserving topology in 3D cardiac image registration tasks. 
\textcolor{edited}{Similarly,} \cite{gao2023diffusing} employed \textcolor{edited}{a diffusion model} \citep{ho2020denoising} to facilitate multimodal \textcolor{edited}{brain \ac{MRI} registration, combining} \ac{DDPM} with a discrete cosine transform module to disentangle structural information, \textcolor{edited}{thus} simplifying the multimodal problem to a quasi-monomodal registration task.

\textcolor{edited}{Overall, existing diffusion-based methods in medical imaging predominantly operate as converter models, translating images from diseased to healthy states or across imaging modalities, rather than as generative models designed to produce anatomically diverse and physically meaningful deformations.}

\subsection{Diffusion model for medical image synthesis and manipulation}

\cite{pinaya2022brain} proposed a 3D \textcolor{edited}{T1-weighted brain} \ac{MRI} synthesis framework based on an \ac{LDM} \citep{rombach2022high}, incorporating a \ac{DDIM} sampler \citep{song2020denoising} to condition the generated images on the subject's age, sex, ventricular volume, and brain volume. 
\textcolor{edited}{Similarly,} \cite{ju2024vm} combined the advanced Mamba network \citep{gu2023mamba} with a cross-scan module into the \ac{DDPM} framework \citep{ho2020denoising} to generate medical images, validated on chest X-rays, brain MRI, and cardiac MRI.

\textcolor{edited}{
Recent advances have further expanded the field of diffusion-based medical image generation.
Fast-DDPM \citep{jiang2025fast} accelerates diffusion-based medical image generation by reducing denoising steps from thousands to only a few while maintaining high image quality and anatomical fidelity. 
DiffBoost \citep{zhang2024diffboost} leverages text-guided diffusion models to synthesize structure-aware medical images, enhancing realism and improving the performance of downstream segmentation tasks. 
}

\textcolor{edited}{
While these appearance-generation methods produce photorealistic images, they face fundamental limitations such as hallucinations and a lack of interpretable correspondence with real anatomical structures \cite{deo2025}.
While some works e.g. Med-DDPM \citep{dorjsembe2024conditional} introduced a conditional diffusion model to generate anatomically consistent 3D MRI synthesis, the loss of correspondence with source images remains unavoidable. Consequently, such generated images cannot be efficiently used for annotation-consistent augmentation for downstream tasks (Sections~\ref{sec:aug_seg} and \ref{sec:syn_reg}).
}

\textcolor{edited}{An alternative strategy is to generate deformation fields rather than image intensities using diffusion modelling.} 
DiffuseMorph \citep{kim2022diffusemorph} uses \ac{DDPM} \citep{ho2020denoising} to estimate the conditional score function for deformation, combined with a deformation module to estimate deformation between image pairs for registration tasks, including 4D temporal medical image generation of cardiac MRI \citep{kim2022diffusemorph}. 
\cite{starck2024diff} introduced a conditional atlas generation framework based on \ac{LDM} \citep{rombach2022high}, generating deformation fields conditioned on specific parameters, with a registration network guiding the optimization of atlas deformation processes.

\textcolor{edited}{
\cite{wu2025igg} formulated image generation as geodesic trajectories within a learned deformation space, enabling anatomically consistent synthesis through continuous shape transformations.
\cite{wang2025generating} proposed a diffusion-based model that learns a population template and generates subject-specific anatomy via deformation-field-driven synthesis of novel 3D brain MRIs.}

\textcolor{edited}{Despite these advances, existing deformation-based diffusion methods still depend on intensity-  \citep{kim2022diffusemorph,kim2022diffusion} or latent-feature-level denoising \citep{starck2024diff,wu2025igg,wang2025generating}, typically within registration-guided frameworks that constrain deformation plausibility}. 
Consequently, the diversity of the generated deformations is largely limited to interpolation between image pairs \textcolor{edited}{\citep{kim2022diffusemorph,kim2022diffusion,wu2025igg}} or the deformation of atlas images \textcolor{edited}{\citep{starck2024diff,wang2025generating}}. \textcolor{edited}{These constraints restrict the diversity of instance-specific deformations and limit their potential for data augmentation or the creation of anatomically diverse deformation fields for downstream tasks.}

\subsection{Data augmentation for few-shot image segmentation}

Several methods have been proposed for few-shot image segmentation, addressing the challenge of limited annotations in medical image analysis \textcolor{edited}{tasks}. \textcolor{edited}{These approaches can be broadly categorised into pseudo-label-based strategies and data augmentation techniques.}

\cite{dang2022vessel} introduced a few-shot learning framework for vessel segmentation, utilising weak and patch-wise annotations. This approach includes synthesising pseudo-labels for a segmentation network and utilising a classifier network to generate additional labels and assess low-quality images. 
\textcolor{edited}{Similarly,} an uncertainty estimation-based mean teacher segmentation method was proposed to enhance the reliable training of a student model in cardiac MRI segmentation \citep{wang2022uncertainty}. 
Another semi-supervised method was introduced based on mutual learning between two Vision Transformers and one Convolutional Network, utilizing a dual feature-learning module and a robust guidance module designed for consistency \citep{wang2022cnn}.

However, pseudo-label-based methods require a sufficient number of annotated labels to ensure the accuracy of an additional model for pseudo-label creation, which limits their applicability in tasks with extremely \textcolor{edited}{low-annotation scenarios}.

\textcolor{edited}{To overcome this limitation}, an atlas-based data augmentation technique \textcolor{edited}{was} introduced in \citep{zhao2019data} to create labelled medical images for brain \ac{MRI} segmentation by spatially and cosmetically aligning an annotated atlas with other images. \textcolor{edited}{However, this method’s diversity is constrained by the variability of the available atlases, and new registration and appearance transformation models must be trained for each atlas.}
\cite{bortsova2019semi} proposed a few-shot segmentation method based on data augmentation through elastic deformation transform \citep{davis1997physics}, with a segmentation consistency loss across labelled and unlabelled images. 
Furthermore, \cite{BigAug} proposed a combination of nine different types of cascaded augmentation methods, named BigAug. These methods vary \textcolor{edited}{in} image quality (sharpness, blurriness, and intensity noise level), image appearance (brightness, contrast, and intensity perturbation), and spatial configuration (rotation, scaling, and deformation), validated on cardiac MRI/ultrasound and prostate MRI. This method, BigAug, has also been compared in Section~\ref{sec:aug_seg}.
\textcolor{edited}{A statistical deformation model combining registration-learned transformations with principal component analysis was proposed by \cite{he2024statistical} for volumetric medical image segmentation. While effective for data augmentation, this approach requires paired images for registration training and generates deformations through interpolation in a learned population space, limiting its capacity for instance-specific deformation synthesis \citep{he2024statistical}.}

\textcolor{edited}{Additionally, \cite{mo2024labelling} proposed a data-efficient approach that learns a vector field from cropped image patches to trace tissue boundaries, achieving strong performance with very limited labeled data on chest X-ray and dermoscopy segmentation. However, it struggles with complex tissue topologies and is difficult to extend to 3D imaging.}

\subsection{Synthetic training for image registration}

\textcolor{edited}{Synthetic spatial transformations have been widely applied to train image registration models, particularly when annotated data are limited or unavailable.} For example, random rigid transformations can be easily synthesised to train a model for rigid registration, aimed at aligning micro-CT scans of murine knees with and without contrast enhancement \citep{zheng2023accurate}.

\cite{rohe2017svf} proposed a training strategy for cardiac MRI deformable registration, based on synthetically deforming the segmented mask of the target tissues via elastic body splines \citep{davis1997physics}. Similarly, \cite{uzunova2017training} adopted a locality-based multi-object statistical shape model method \citep{wilms2017multi} for statistical appearance modelling, to synthesise training data for medical image registration \citep{uzunova2017training}. 
However, these methods rely on segmentation masks or statistical shapes of the target tissues prior to training the registration model, making them unsuitable for unsupervised training approaches.

\textcolor{edited}{eliminate the need for} annotations, random deformations based on Gaussian smoothing sampling \citep{sokooti2017nonrigid,gonzales2021moconet} have been used for registration model training. Recently, a mixture of Gaussian and thin-plate splines \citep{zheng2022recursive, zheng2022residual} have also been used for pretraining registration models. 
\textcolor{edited}{For pulmonary \ac{CT} registration tasks, \citet{eppenhof2018pulmonary} applied a multi-resolution B-spline~\citep{lee1997scattered} model to generate random deformations for data synthesis and model pretraining. This B-spline-based method serves as a key comparison baseline for synthetic registration experiments in Section~\ref{sec:syn_reg}.}



\section{Discussion and Conclusion }
\label{sec:discuss_conclude}
\subsection{Plausible and diverse deformation synthesis}

The experiments conducted on cardiac \ac{MRI} and pulmonary \ac{CT}, as \textcolor{edited}{presented} in Section~\ref{sec:img_syn}, demonstrate that \textcolor{edited}{the proposed} method, \ac{DRDM}, \textcolor{edited}{is capable of generating anatomically plausible and diverse deformations for instance-specific images}. Unlike previous deformation methods \citep{kim2022diffusemorph,kim2022diffusion,starck2024diff}, \ac{DRDM} does not \textcolor{edited}{rely on} a registration framework to guide the deformation process. \textcolor{edited}{This independence enables the model to produce a broader range of deformation patterns while maintaining anatomical coherence.} 
In comparison to earlier deformation-based augmentation methods \citep{BigAug,eppenhof2018pulmonary}, \ac{DRDM} \textcolor{edited}{generates more customized and anatomically consistent} deformations for each individual image.

\textcolor{edited}{The apparent diversity of deformations in 2D cardiac images is inherently more limited than in 3D volumes. This restriction arises because 2D slices capture fixed cross-sections of a 3D anatomy, with fewer degrees of freedom for spatial variation. Simulating inter-slice transformations that mimic positional changes across the 3D anatomy would introduce discontinuities, causing tissue to appear or disappear between slices, which could compromise anatomical correspondence and degrade downstream task performance.}

\subsection{Generated artifacts and unreasonable structures}

\begin{figure}
    \centering
    \includegraphics[width=1.\linewidth]{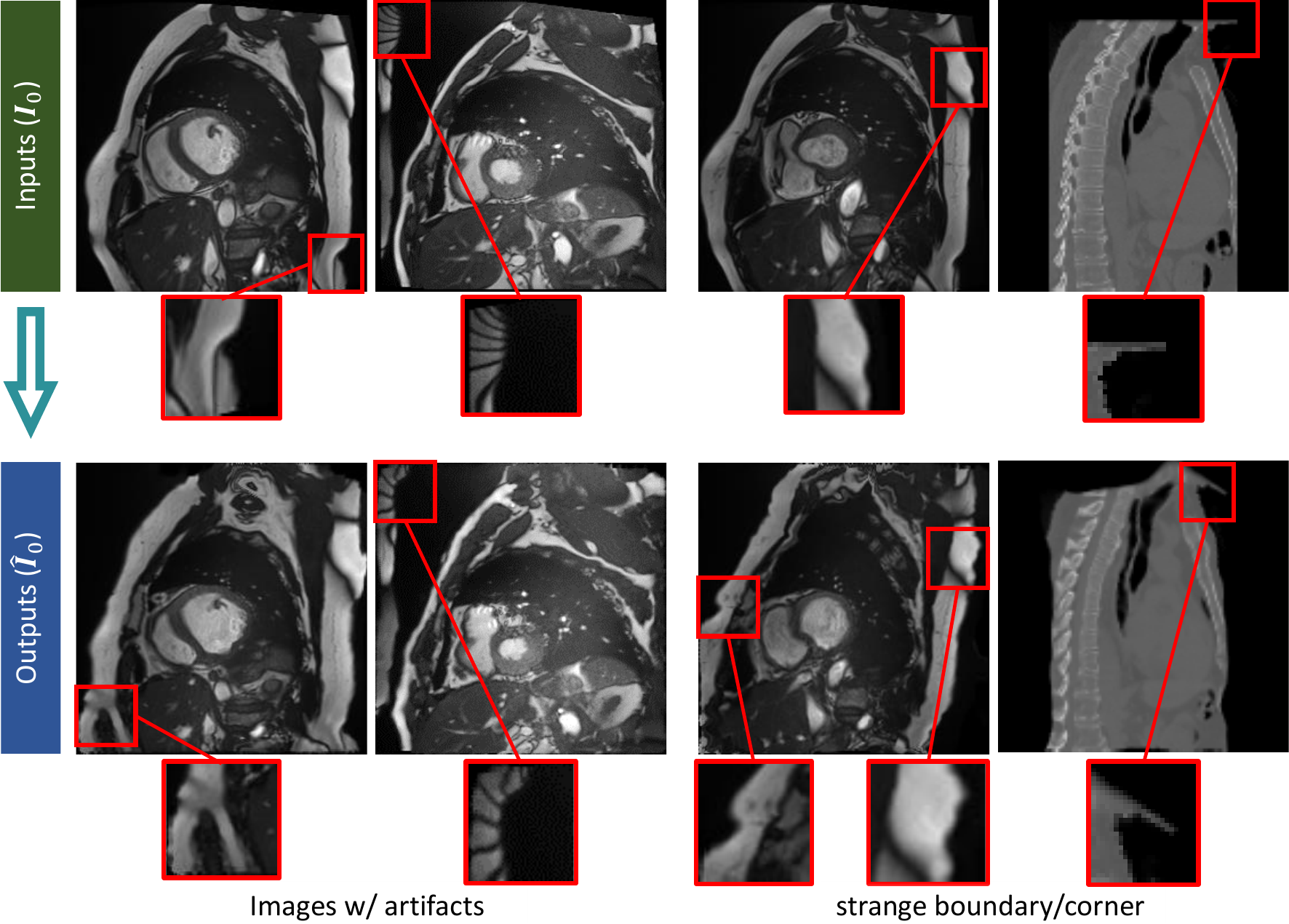}
    \caption{\textcolor{edited}{Artifacts from the original MRI and To occasional cropping of unexpected tissue in CT To images can result in atypical or distorted structures in the To generated outputs (added according to Comment-1.3)}}
    \label{fig:artifacts}
\end{figure}




\textcolor{edited}{
We observed artifacts characterized by sharp corners and irregular boundaries in some generated images (Figure~\ref{fig:artifacts}). Further investigation identified two primary sources of these anomalies: inherent artifacts present in the original \ac{MRI} acquisitions, and instances where anatomical boundaries were inadvertently cropped in the original \ac{CT} data. 
}

\textcolor{edited}{
Although such imperfections may be undesirable when artifact-free images are required, they can enhance the realism for data augmentation by preserving the noise characteristics of clinical acquisitions. Importantly, the negative Jacobian determinant ratio (Table~\ref{tab:def_mag_detj}) remains below 1\% even for large deformations, confirming that the generated deformation fields are smooth and topologically valid.
}

\subsection{Improvement of downstream task}

As described in Section~\ref{sec:aug_seg} and Section~\ref{sec:syn_reg}, \textcolor{edited}{the additional} experiments on segmentation and registration tasks \textcolor{edited}{confirm} the efficacy and applicability of the diverse and instance-specific deformations generated by \ac{DRDM}. Previous augmentation methods typically rely on fully random transformations in image quality, image appearance, and spatial features, without \textcolor{edited}{adapting to the characteristics of} each individual image. In contrast, \ac{DRDM} synthesises more realistic images and thus improves the downstream-task models in learning the realistic distribution of images by balancing diversity and plausibility. 
It is also noteworthy that \ac{DRDM} can be combined with other data augmentation methods to further enhance downstream tasks. \textcolor{edited}{The observed improvements in segmentation and registration tasks collectively validate the reliability, generalizability, and effectiveness of the deformations generated by \ac{DRDM}.}

\subsection{Limitations of this research}

\textcolor{edited}{The objective of this study} is to generate diverse, high-quality, and realistic image deformations. Although experimental results in Section~\ref{sec:img_syn} show that the generated deformations are diverse and reasonable, \textcolor{edited}{the perceptual realism of these deformations can only be evaluated qualitatively. 
Quantitative assessment of visual realism remains challenging, a limitation common to image generation research \cite{deo2025}}.

\textcolor{edited}{The Fréchet Inception Distance (FID) is widely used to measure the similarity between generated and real image distributions in natural image synthesis~\citep{rombach2022high,ho2020denoising,song2020denoising,qiao2019mirrorgan}. However, applying FID to medical imaging tasks is problematic due to domain-specific factors, including misaligned data distributions~\citep{jayasumana2024rethinking} and the lack of suitable pretrained feature extractors for medical images.}

Therefore, \textcolor{edited}{in this study, the evaluation of \ac{DRDM} relies on its effectiveness} in downstream tasks, such as segmentation and registration, to demonstrate the quality and clinical utility of the synthesised medical images. The improvement \textcolor{edited}{observed in these} tasks can prove that the synthetic images and deformation generated by \ac{DRDM} \textcolor{edited}{align} to the data distributions learned by the segmentation and registration models in the downstream tasks. 
\textcolor{edited}{Nevertheless, it should be noted that the distributions captured by these task-specific models are only approximations of real-world data distributions, and further investigation is required to establish a direct quantitative link.}

\subsection{Prospective applications in future}

This paper demonstrates the application value of \ac{DRDM} for data augmentation in few-shot segmentation and data synthesis for registration. There is considerable potential for exploring other directions. 
\textcolor{edited}{The \ac{DRDM} can be modified to accept conditional inputs that regulate the generated deformation fields and their corresponding deformed images for specific applications, including conditional image registration and text-guided image synthesis.} A segmentation module can be employed to decompose different regions of images, enabling \ac{DRDM} to generate more complex deformation fields with multiple continuums. An image modality converter module can be combined to generate deformed images in another modality. 
\textcolor{edited}{
Furthermore, \ac{DRDM} can be combined with conventional intensity-based or latent-based diffusion models to address a broader range of variations. Such hybrid approaches would use \ac{DRDM} for anatomical deformations while intensity-based models handle appearance changes, textural variations, and pathological features. This combination could be particularly valuable for generating pathological variations, addressing data imbalance in rare conditions, and creating realistic temporal sequences with both anatomical motion and appearance changes.
} 

\subsection{Conclusion}

In this \textcolor{edited}{study}, we proposed a novel diffusion-based deformation generative model, \textcolor{edited}{termed} \ac{DRDM}, for image manipulation and synthesis \textcolor{edited}{in medical imaging}. The experimental results \textcolor{edited}{demonstrate} that \ac{DRDM} achieves \textcolor{edited}{both anatomical plausibility and diversity in the generated deformations and substantially enhances the performance of downstream tasks, including cardiac MRI segmentation and pulmonary CT registration}.
\textcolor{edited}{These findings highlight the potential of \ac{DRDM} as a general framework for anatomically consistent image synthesis, deformation modelling, and data augmentation in a wide range of medical imaging applications.}

\section{Acknowledgements}
J.-Q. Z. acknowledges the Kennedy Trust Prize Studentship (AZT00050-AZ04) and the Chinese Academy of Medical Sciences (CAMS) Innovation Fund for Medical Science (CIFMS), China (grant number: 2018-I2M-2-002).
B.W.P. acknowledges the Rutherford Fund at Health Data Research UK (grant no. MR/S004092/1).

\section{Declaration of AI technologies used in writing}
During the preparation of this work the authors used ChatGPT\footnote{OpenAI. (2024). ChatGPT (4o) [Large language model]. https://chatgpt.com} in order to proofread the text. After using this tool/service, the author(s) reviewed and edited the content as needed and take(s) full responsibility for the content of the published article.


\appendix

\begin{table}[thb]
\caption{{Network structure detail for DRDM.}}
\centering
\begin{tabular}{ccccc}
\toprule
\multirow{2}{*}{func}&\multirow{1}{*}{spatial}&\multirow{1}{*}{\#chnl}&\multirow{2}{*}{in}&\multirow{2}{*}{out}\\
\cline{3-3}
&size&in/out&&
\\
\midrule\midrule
embed&1,1,1&1/80&$t$&t0\\
fc,act,fc&{1,1,1}&80/1&t0&t1\\
fc,act,fc&{1,1,1}&80/10&t0&t2\\
fc,act,fc&{1,1,1}&80/20&t0&t3\\
fc,act,fc&{1,1,1}&80/40&t0&t4\\
fc,act,fc&{1,1,1}&80/80&t0&t5\\
fc,act,fc&{1,1,1}&80/40&t0&t6\\
fc,act,fc&{1,1,1}&80/20&t0&t7\\
\midrule
ACNN&H,W,D&$c_0$/10&$\hat{\textbf{\textit{I}}}_t$+t1&f1\\
ACNN&H,W,D&10/10&f1&f1\\
ACNN&H,W,D&10/10&f1&f1\\
\midrule
stride conv&H/2,W/2,D/2&10/10&f1&f2\\
\midrule
ACNN&H/2,W/2,D/2&10/20&f2+t2&f2\\
ACNN&H/2,W/2,D/2&20/20&f2&f2\\
ACNN&H/2,W/2,D/2&20/20&f2&f2\\
\midrule
stride conv&H/4,W/4,D/4&20/20&f2&f3\\
\midrule
ACNN&H/4,W/4,D/4&20/40&f3+t3&f3\\
ACNN&H/4,W/4,D/4&40/40&f3&f3\\
ACNN&H/4,W/4,D/4&40/40&f3&f3\\
\midrule
stride conv&H/8,W/8,D/8&40/40&f3&f4\\
\midrule
ACNN&H/8,W/8,D/8&40/20&f4$\times$t4&f4\\
ACNN&H/8,W/8,D/8&20/20&f4&f4\\
ACNN&H/8,W/8,D/8&20/40&f4&f4\\
\midrule
trans conv&H/4,W/4,D/4&40/40&f4&f5\\
\midrule
ACNN&H/4,W/4,D/4&80/40&f5$|$f3+t5&f5\\
ACNN&H/4,W/4,D/4&40/20&f5&f5\\
ACNN&H/4,W/4,D/4&20/20&f5&f5\\
\midrule
trans conv&H/2,W/2,D/2&20/20&f5&f6\\
\midrule
ACNN&H/2,W/2,D/2&40/20&f6$|$f2+t6&f6\\
ACNN&H/2,W/2,D/2&20/10&f6&f6\\
ACNN&H/2,W/2,D/2&10/10&f6&f6\\
\midrule
trans conv&H,W,D&10/10&f6&f7\\
\midrule
ACNN&H,W,D&20/10&f7$|$f1+t7&f7\\
ACNN&H,W,D&10/10&f7&f7\\
ACNN&H,W,D&10/10&f7&f7\\
\midrule
conv &H,W,D&10/3&f7&$\hat{\psi}_t^{(k)}$\\
\bottomrule
\label{tab:net_drdm}
\end{tabular}
\end{table}

\begin{table}[thb]
\caption{Network structure detail for the $i^{\rm th}$ ACNN-II block.}
\centering
\begin{tabular}{ccccc} 
\toprule
\multirow{2}{*}{func}&\multicolumn{1}{c}{kern param}&\multirow{1}{*}{\#chnl}&\multirow{2}{*}{in}&\multirow{2}{*}{out}\\
\cmidrule(lr){2-2}\cmidrule(lr){3-3}
&dila/str/pad&in/out&&
\\
\midrule\midrule
conv,norm & 1/1/1&$c_i$/$c_{i+1}$
&fi&r\\
act,conv& 1/1/1  &$c_{i+1}$/$c_{i+1}$
&r&fo\\
act,conv& 3/1/3  &$c_{i+1}$/$c_{i+1}$
&fo&fo\\
act& -  &$c_{i+1}$/$c_{i+1}$
&(fo+r)&fo\\
\bottomrule
\label{tab:net_acnn}
\end{tabular}
\end{table}

\section{Network architecture for {DRDM}}
\label{sec:net_detail_drdm}
The network structure detail for the \ac{DRDM} is shown in Table~\ref{tab:net_drdm}, where "embed" denotes feature embedding, "fc" denotes a fully connected layer, "act" denotes the \ac{ReLU} activation function with 0.01 negative slope, "\#chnl" denotes channel number for input or output, "conv" denotes the convolution with kernel size of 3, stride of 1 and padding size of 1, "stride conv" denotes convolution with stride of 2, "trans conv" denotes transpose convolution, and "ACNN" denotes the ACNN-II block \citep{zhou2020acnn} as described in Table~\ref{tab:net_acnn}. As described in Equation~\eqref{eq:net_est_dvf}, one image $\hat{\textbf{\textit{I}}}_t$ is fed into \ac{DRDM} network and a \ac{DVF} $\hat{\psi}_t^{(k)}$ is predicted.

As shown in Table~\ref{tab:net_acnn}, the network structure detail for the $i^{\rm th}$ ACNN-II block, where  "conv" denotes convolution, "ker param" denotes kernel parameters, with "dila" as dilation rate, "str" as the stride rate, and "pad" as the padding size, "norm" denotes the instance normalization, and "act" denotes the leaky \ac{ReLU} activation function with $10^{-6}$ negative slope. The input of ACNN is a feature map with $c_i$ and the output is the feature map with $c_{i+1}$ processed by three convolution or dilated convolution and activation function.

\begin{figure}[!ht]
\begin{center}
\includegraphics[width=1.\linewidth]{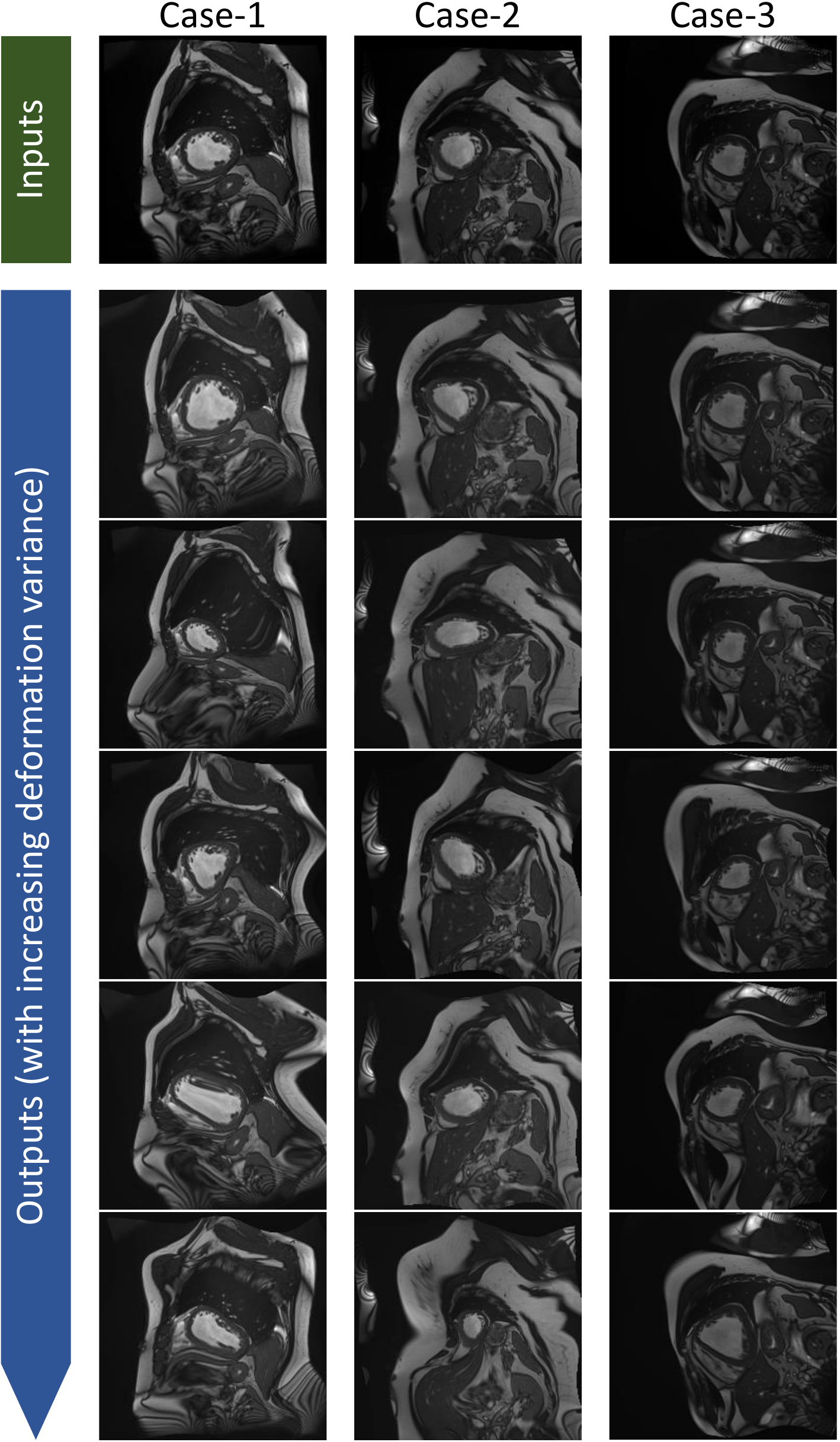}
\end{center}
\vspace{-.5em}
\caption{The original and deformed images of three subjects by Elastic transformation as used in BigAug for 2D cardiac MRI scans.}
\label{fig:examples_2d_base}
\end{figure}

\begin{figure}[!ht]
\begin{center}
\includegraphics[width=1.\linewidth]{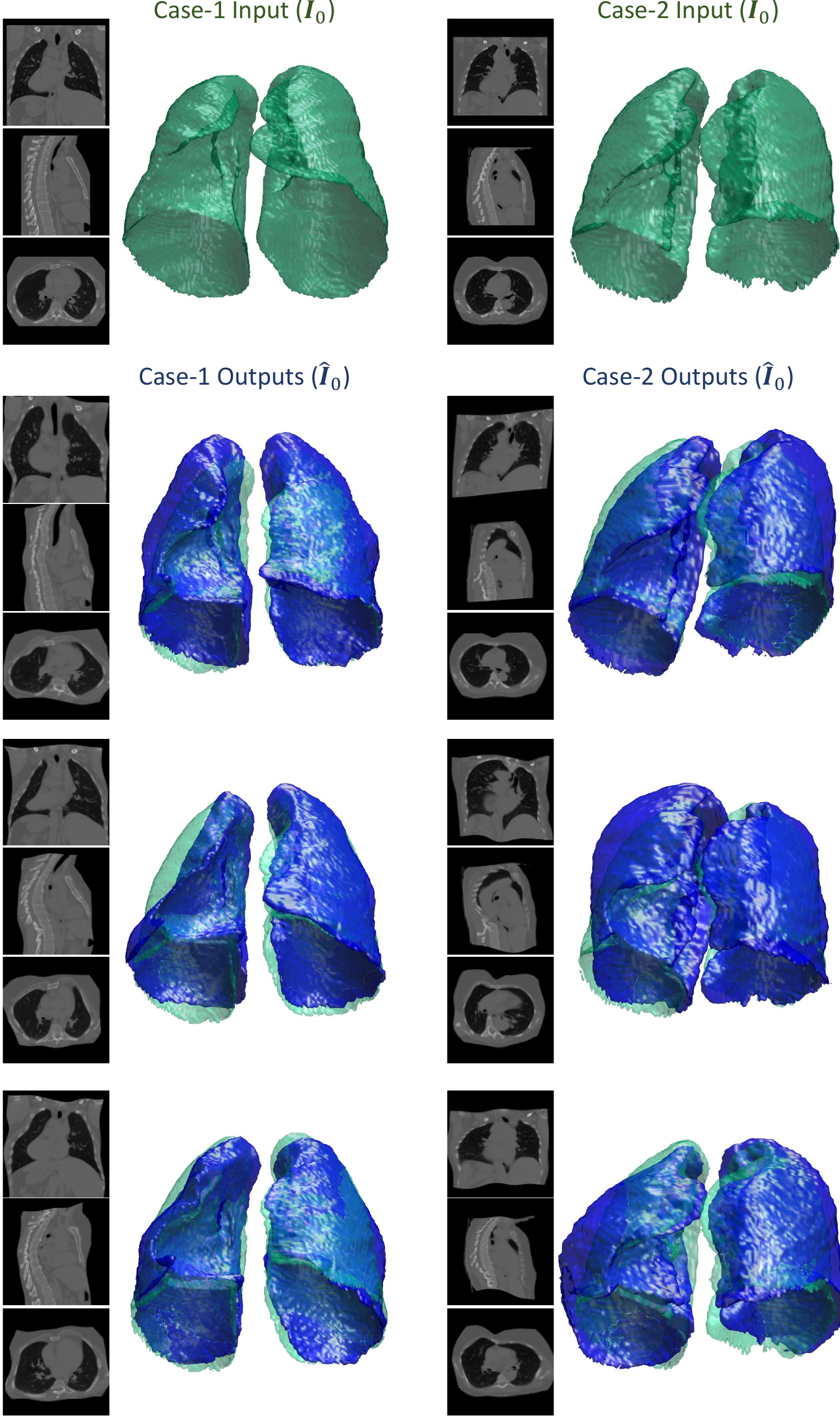}
\end{center}
\vspace{-.5em}
\caption{The original and deformed images of two subjects by MRBS for 3D pulmonary CT scans.}
\label{fig:examples_3d_base}
\end{figure}

\section{Baseline results for image deformation}
\label{sec:def_base}

The baseline results for image deformations are illustrated in Figure~\ref{fig:examples_2d_base} using Elastic transform (a part of BigAug) \citep{BigAug} and Figure~\ref{fig:examples_2d_base} using \ac{MRBS} \citep{eppenhof2018pulmonary}. Notably, the deformations appear unrealistic, such as the unnaturally expanded or squeezed ventricles and the distorted body shape shown in Figure~\ref{fig:examples_2d_base} and the unnatural shearing lung in Figure~\ref{fig:examples_3d_base}. These unrealistic deformations can negatively impact the effectiveness of data augmentation or data synthesis, as validated by the experimental results in Section~\ref{sec:aug_seg} and Section~\ref{sec:syn_reg}.


%
%
\clearpage

\bibliographystyle{model2-names.bst}\biboptions{authoryear}
\bibliography{refs}


\end{document}